\documentclass[11pt,
amsmath,
amssymb,
amsfonts,
superscriptaddress,
tightenlines,
nobibnotes,
prd,
nofootinbib,
twocolumn,
dvipsnames]{revtex4-2}

\usepackage{bm}
\usepackage[utf8]{inputenc}
\usepackage{amsmath}
\usepackage{amssymb}
\usepackage{xspace} %for handling spaces after macros
\usepackage{subeqnarray}
\usepackage{graphicx}
\usepackage[caption=false]{subfig} %allows subcaptions in figures
\usepackage[table]{xcolor}
\usepackage{tikz}
\usetikzlibrary{arrows.meta, positioning, calc, decorations.pathmorphing}

\tikzset{
  block/.style = {
    rectangle,
    draw=green!60!black,
    fill=green!20,
    text width=4cm,
    align=center,
    minimum height=1.2cm,
    font=\sffamily
  },
  redblock/.style = {
    block,
    draw=red!70!black,
    fill=red!20
  },
  yellowblock/.style = {
    block,
    draw=yellow!50!black,
    fill=yellow!20
  },
  blueblock/.style = {
    block,
    draw=blue!70!black,
    fill=blue!20
  },
  arrow/.style = {
    draw,
    -{Latex[length=2mm]},
    thick
  },
  curvedarrow/.style = {
    arrow,
    bend left=30
  }
}

\usepackage{yfonts}
\usepackage{orcidlink}
\usepackage{epsfig}
\hypersetup{
     hidelinks,
     colorlinks=true,
     citecolor=Blue
 } 
\usepackage{cleveref}
\usepackage{booktabs}
\usepackage{soul}

\newcommand\spart{\;\raise1.0pt\hbox{/}\hskip-6pt\partial}
\newcommand\spartb{\;\overline{\raise1.0pt\hbox{/}\hskip-6pt\partial}}

\newcommand{\be}{\begin{equation}}
\newcommand{\ee}{\end{equation}}
\newcommand{\bea}{\begin{eqnarray}}
\newcommand{\eea}{\end{eqnarray}}
\newcommand{\beal}{\begin{align}}
\newcommand{\eeal}{\end{align}}
\newcommand{\beas}{\begin{subeqnarray}}
\newcommand{\eeas}{\end{subeqnarray}}
\newcommand{\dd}{{\rm d}}
\newcommand{\ii}{{\rm i}}
\newcommand{\cH}{{\cal H}}

\newcommand{\cS}{{\cal S}}
\newcommand{\cN}{{\cal N}}
\newcommand{\cG}{{\cal G}}
\newcommand{\mykp}[3]{{}_{#1}\kappa^{#2}_{#3}}
\newcommand{\myC}[3]{{}^{#1}{\cal C}^{#2}_{#3}}
\newcommand{\ellc}{\ell_c}
\newcommand{\ells}{\ell_s}

\newcommand{\class}{\texttt{CLASS}\xspace}
\newcommand{\aniclass}{\texttt{AniCLASS}\xspace}
\newcommand{\anilos}{\texttt{AniLoS}\xspace}

\newcommand{\tspcolor}[1]{\textcolor{black}{#1}}
\newcommand{\tsp}[1]{\textcolor{black}{#1}}

\newcommand{\python}{\texttt{Python}\xspace}
\newcommand{\C}{\texttt{C}\xspace}

\begin{document}

\title{CMB line-of-sight integrators for nearly-isotropic cosmological models}

\author{João G. Vicente~\orcidlink{0009-0008-2912-8995}}
\email[]{jgabvicente.2000@uel.br }
\affiliation{Departamento de Física, Universidade Estadual de Londrina, Rod. Celso Garcia Cid, Km 380, 86057-970, Londrina, Paraná, Brazil.}

\author{Thiago S. Pereira~\orcidlink{0000-0002-6479-364X}}
\email[]{tspereira@uel.br}
\affiliation{Departamento de Física, Universidade Estadual de Londrina, Rod. Celso Garcia Cid, Km 380, 86057-970, Londrina, Paraná, Brazil.}
\affiliation{Instituto de Física, Universidade Federal do Rio de Janeiro, 21941-972, Rio de Janeiro, RJ, Brazil.}

\author{Cyril Pitrou~\orcidlink{0000-0002-1747-7847}}
\email[]{pitrou@iap.fr}
\affiliation{Institut d'Astrophysique de Paris, CNRS UMR 7095, 98 bis Bd Arago, 75014 Paris, France.}

%================================================
\begin{abstract}
%================================================
Homogeneous and nearly-isotropic cosmological models are natural extensions of standard Friedmann cosmologies. Constraining their features is crucial, as any detection of their properties would impact our understanding of inflation and the cosmological principle. Since these models evolve as a set of non-interacting scalar, vector, and tensor modes on top of homogeneous and isotropic spacetimes, their imprints on cosmological observables, particularly the CMB, can be obtained using standard line-of-sight methods. This requires (1) that one resorts on Laplacian eigenmodes on spatially curved spaces and (2) that radial functions for these modes are analytically continued to accommodate complex (i.e., supercurvature) wavenumbers. We introduce two line-of-sight integrators implementing the evolution of the CMB anisotropies in these models: \anilos, a user-friendly and easy to modify \python package, and \aniclass, an advanced and efficient extension of the Boltzmann solver \class. We discuss possible initial conditions that could generate such fluctuations and provide illustrative examples using our codes. This work offers a pathway for leveraging diverse cosmological datasets to constrain superhorizon anisotropies of the late-time universe.
\end{abstract}

\maketitle

%
%================================================
\section{Introduction}
%================================================

A long-standing cosmological question concerns the global shape of the universe. According to the inflationary paradigm, our particle horizon is just a smooth patch of a larger manifold whose exact geometrical and topological properties are unknown. The extension of this patch beyond our particle horizon and the details of the underlying spacetime manifold are uncertain, and both have profound implications for inflation and the cosmological principle. At the same time, the existence of large-angle statistical anomalies in the cosmic microwave background (CMB) and persistent tensions between low- and high-redshift data \cite{Aluri:2022hzs,Abdalla:2022yfr}, suggest that new physics may be lurking just beyond the horizon. These considerations motivate the search for observational signatures that could reveal the global structure of spacetime while also testing the robustness of the concordance Friedmann-Lemaître-Robertson-Walker (FLRW) model. This program requires relaxing some or all of the foundational symmetry assumptions of our universe, namely, that it is spatially homogeneous and isotropic at large scales \tspcolor{(see \cite{Krasinski:1997yxj,ellis2012relativistic} for classical references and \cite{Buchert:2001sa,Heinesen:2022lqs} for recent accounts), and with topologically trivial boundary conditions \cite{Lachieze-Rey:1995qrb}}. While past searches for nontrivial topologies resulted inconclusive~\cite{Planck:2013okc,Planck:2015gmu,Luminet:2016bqv}, this is an active area of research that might be settled with existing and future cosmological data~\cite{COMPACT:2022gbl}. Another possibility is to keep the trivial topology, but change the geometrical setup. In this case the simplest option is to describe the universe using homogeneous and spatially anisotropic Bianchi models.

Among all Bianchi models, those built out of maximally symmetric three-dimensional spaces are phenomenologically interesting, due to their FLRW limit of zero anisotropy. This includes the Euclidean models I and VII$_0$, the hyperbolic models V and VII$_h$, and the spherical model IX. When their anisotropies are perturbatively small, these models are usually called \emph{nearly-isotropic}.

Nearly-isotropic models possess two independent classes of solutions~\cite{Pontzen:2009rx}. The first class includes solutions diverging at the Big Bang singularity, and are called irregular. Regular solutions, on the other hand, are finite at the singularity, and can develop nontrivial dynamics at later times. It is thus natural to seek constraints for these models using both high- and low-redshift cosmological data.

At very high redshifts ($z\sim10^{11}$), Big Bang nucleosynthesis (BBN) provides the strongest observational constraints available. This follows since any deviation of the expansion rate from isotropy will alter {\it i}) the time at which weakly interacting particles decouple and {\it ii}) the time elapsed between this decoupling and the onset of nucleosynthesis, hence affecting the amount of neutrons lost by beta decay. The modified neutron-proton ratio abundance when nuclear reactions take place then affect the deuterium and helium-4 mass fractions~\cite{Pitrou:2018cgg}. Since measurements of these quantities are in excellent agreement with FLRW predictions, they translate into strong limits to the simplest anisotropic models \cite{Thorne:1967zz,Barrow:1976rda,barrow1977homogeneity,olson1978helium}, though more sophisticate models are still consistent with BBN data~\cite{barrow1984helium,Pontzen:2009rx}.

At redshifts $z\sim10^3$, cosmic microwave background (CMB) anisotropies are arguably the best observational window to constrain large scale deviations from isotropy. In fact, since CMB probes the universe at horizon-scales, it allows us to test the isotropy hypothesis at the onset of inflation~\cite{Pitrou:2008gk}. To date, the tightest and most comprehensive constraints on large scale anisotropies were obtained with Planck's temperature and polarization data in Ref.~\cite{Saadeh:2016sak}, where it was shown that the anisotropic expansion rate produced by vectors modes are $(\sigma_V/H)_0\lesssim10^{-10}$, and thus largely consistent with zero. Tensor modes were found to produce much weaker constraints, $(\sigma_T/H)_0 < 10^{-6}$~\cite{Saadeh:2016sak}, though this will certainly improve with the next generation of CMB polarization experiments.

At intermediate and low redshifts, there are many observables available to test isotropy, such as luminosity distances~\cite{Campanelli:2010zx,Javanmardi:2015sfa,Bengaly:2015dza,Zhao:2019azy}, cosmological drifts~\cite{Quercellini:2009ni,Marcori:2018cwn}, and gravitational weak-lensing \cite{Adam:2024kgs,Pereira:2015jya,Pitrou:2015iya}, to name a few. For example, a measurement of $E$- and $B$-modes of the weak-lensing shear by Euclid can be translated in constraints of the order $(\sigma/H_0)\leq1\%$~\cite{Pereira:2015jya}. While these are not as stringent as CMB or BBN constraints, they are complementary, since the physics leading to late-type anisotropy can in principle evade high-redshift constraints.

In principle, any cosmological observable in a given nearly-isotropic model can be computed and compared against observations. In practice, however, this is a much more involved task than its isotropic analogs, and its success depends strongly on the type of observable chosen. Nonetheless, if spatial anisotropies are small, as observations suggest, one can show that the anisotropies will evolve as an independent set of modes on top of a maximally symmetric space---just as it happens with the usual scalar, vector and tensor modes of linear perturbation theory. This idea was originally introduced in \cite{chiu1964gravitation,Grishchuk:1975ec,King:1991jd}, where homogeneous tensor perturbations of spherical FLRW universes were identified with the dynamics of Bianchi IX models. It was then generalized in \cite{Pontzen:2010eg} to include the proper definition of scalar, vector and tensor modes of nearly-isotropic models, and their identification with the dynamics of perturbative modes of FLRW spacetimes.

More recently, it was shown that, starting from linear perturbation theory in synchronous gauge, there exists a long-wavelength (i.e., homogeneous) limit where FLRW perturbations are dynamically equivalent to the Bianchi perturbations~\cite{Pereira:2019mpp}. This result has an important practical implication: provided that the anisotropies are small, the expression for a cosmological observer in a nearly-isotropic universe can be obtained by computing that same observer in a perturbed FLRW universe, and then taking the long-wavelength limit of the resulting expression. This expedient has been used to compute weak-lensing \cite{Adamek:2015mna} and cosmological-drifts \cite{Marcori:2018cwn} observables in Bianchi I spacetimes.

In the simplest case of a Bianchi I universe, where the identification is made with perturbations of a flat FLRW metric, the long-wavelength limit corresponds to taking $k\rightarrow0$, where $k$ is the perturbation's wavenumber. However, for other Bianchi spacetimes, including those where the correspondence is made with perturbations of curved FLRW geometries, the proper definition of the long-wavelength limit is not trivial, and allows for the presence of supercurvature modes, i.e., modes whose wavelength is larger than the curvature radius of the universe~\cite{Pereira:2019mpp,Lyth:1995cw}.

In this work we complete a program started in \cite{Pitrou:2019ifq} and \cite{Pereira:2019mpp}, and show that, given a set of initial conditions and cosmological parameters, the CMB signatures of nearly-isotropic models can be extracted from standard line-of-sight integration techniques, provided that the appropriate generalization of plane-waves and spherical Bessel functions to curved spaces \cite{Hu:1997mn}---accounting in particular for the possibility of supercurvature modes---is numerically implemented. To this end, we introduce two numerical line-of-sight integrators for homogeneous cosmological perturbations: \anilos and \aniclass. While they execute the same tasks, they target different use cases. \anilos (Anisotropic Line-of-Sight) is a user-friendly \python package that is easy to read and modify. \aniclass (Anisotropic \class) is a branch of the popular \class code \cite{Blas:2011rf}, inheriting several advanced features from it, and targeted at performance. Both codes are freely available at \cite{gitjoao}.

The details of the formal correspondence between nearly-isotropic models and homogeneous FLRW perturbations are highly technical, and it is not our purpose to rederive them here. Instead, we give in Section~\ref{sec:formalism} a qualitative exposition of the main ideas, leaving the details to \cite{Pereira:2019mpp} and the references therein. In Section~\ref{sec:CMB_and_los} we explain how CMB anisotropies in nearly-isotropic models can be derived from the dynamics of linear perturbations in FLRW universes, emphasizing the necessary modifications in standard line-of-sight integrators to achieve this task. In Section~\ref{sec:anilos_and_aniclass} we explain how \anilos and \aniclass are organized. We compare the two codes and give some examples of their use in Section~\ref{sec:examples}. We conclude in Section~\ref{sec:conclusions}.

%================================================
\section{Homogeneous FLRW perturbations as Bianchi models}\label{sec:formalism}
%================================================
The connection between homogeneous FLRW cosmological perturbations and Bianchi models arises from a simple observation: in standard perturbation theory, spacetime is assumed to possess maximal spatial symmetry. Perturbations are then introduced as small fluctuations that break these symmetries. However, since these perturbations are arbitrary (apart from having small amplitudes), one may ask whether a subset of the original symmetries can be recovered in a limit where perturbations become invariant under a subgroup of the initial symmetry group. In the particular case at hand, we are looking for a limit where the inhomogeneous metric perturbations of FLRW spacetimes in synchronous gauge
\begin{equation}\label{eq:metric-flrw}
\boldsymbol{g}^\text{FLRW} = -\dd t^2 + a^2(t)[\gamma_{ij} + 2h_{ij}]\dd x^i\otimes\dd x^j\,,
\end{equation}
can be identified with the homogeneous metric perturbations of nearly-isotropic Bianchi models,
\begin{equation}\label{eq:metric-small-beta}
 \boldsymbol{g}^\text{Bianchi} = -\dd t^2 + a^2(t)[\delta_{ij} + 2\beta_{ij}]\boldsymbol{e}^i\otimes\boldsymbol{e}^j\,.
\end{equation}
In other words, we look for the identification
\begin{equation}\label{eq:duality}
 h_{ij}\;\,{\longrightarrow}\;\,\beta_{ij}
\end{equation}
in a suitably-defined homogeneous limit.

Naturally, this identification comes with important caveats. The first and most obvious is related to the choice of basis, since the tensor $h_{ij}$ is usually implemented in a coordinate basis $\{\dd x^i\}$, whereas $\beta_{ij}$ is better described in terms of the so-called invariant basis $\{\boldsymbol{e}_i\}$, defined so as to respect the Bianchi symmetries. In particular, this distinction means that the homogeneous limit does not necessarily imply a homogeneous (i.e., position-independent) $h_{ij}$~\cite{Pontzen:2010eg}. Second, this identification holds for Bianchi models having maximally symmetric spaces as their isotropic limits. This includes models I and VII$_0$ (having $\mathbb{E}^3$ as limit), models V and VII$_h$ ($\mathbb{H}^3$), and model IX ($\mathbb{S}^3$). \tsp{Models with anisotropic spatial geometries, also known as Thurston's geometries~\cite{Thurston:1982zz}, do not fall in this class, although they can also be constrained by CMB~\cite{Pereira:2015pxa,Smith:2024map} and distance measures~\cite{Awwad:2022uoz}.} Finally, since $\beta_{ij}$ is traceless (because it is a volume-preserving deformation tensor), the trace of $h_{ij}$ does not play a role in the identification.\footnote{The trace of $h_{ij}$ corresponds to a local curvature of the space in the homogeneous limit, and can be absorbed in a redefinition of the scale factor in Friedmann's equation to account for this curvature --- see~\cite{Norena:2024miy}.}

That this identification is possible can be informally illustrated with the example of gravitational waves in flat FLRW spaces, where the homogeneous limit is trivially reached by making perturbations independent of position. Computing the linearized Einstein equations for a symmetric, traceless and transverse tensor $h_{ij}$, we find in this limit the well-known gravitational wave equation
\begin{equation}
 h''_{ij} + 2\cH h'_{ij} = 0\,,
\end{equation}
where $\cH$ is the conformal Hubble factor and a prime means derivative with respect to $\eta = \int \dd t/a$. This is dynamically equivalent to the equation for the symmetric and traceless shear tensor\footnote{We caution the reader that we are reserving the term \emph{shear} for the metric perturbation $\beta_{ij}$, while in most references this term is associated with the tensor ${\sigma_{ij}\equiv\beta'_{ij}}$.} $\beta_{ij}$ in a Bianchi-I spacetime:
\begin{equation}
\beta''_{ij}+2\cH\beta'_{ij}=0\,,                                                                                                                                                       \end{equation}
as one can easily check by computing the tracefree part of Einstein equations from~\eqref{eq:metric-small-beta} using $\boldsymbol{e}^i = \dd x^i$. An important difference in the two approaches is related to the choice of initial conditions. While in FLRW we are interested in the growing mode of $h_{ij}$, the growing mode of $\beta_{ij}$ is constant in this case, and thus a gauge artifact. Since decaying modes of $\beta_{ij}$ usually diverge towards the Big Bang, some care is needed to chose a non-trivial growing mode of the shear, as we will see.

In the general case, the identification \eqref{eq:duality} is done in terms of the modes of the perturbations that transform as scalars, vectors, or tensors under spatial rotations. For $h_{ij}$, it is usual to consider rotations around a fixed Fourier vector $\boldsymbol{\nu}$ when defining these modes, which leads to well-known scalar-vector-tensor (SVT) decomposition\footnote{For closed FLRW universes, eigenfunctions are periodic, so that $\boldsymbol{\nu}=(\nu,\ell,M)$ is discrete, and the integral is replaced by a sum.}
\begin{equation}\label{eq:SVT}
h_{ij} = \sum_{m=-2}^2\int\frac{\dd^3\boldsymbol{\nu}}{(2\pi)^3}h_{(m)}(\boldsymbol{\nu},\eta)Q^{(m)}_{ij}(\boldsymbol{\nu})\,,
\end{equation}
where $Q^{(m)}_{ij}$ are symmetric and traceless tensor harmonics of maximally symmetric 3-spaces [see Ref. \cite{Pitrou:2019ifq} for their derivation] and $m=0$, $\pm1$, and $\pm2$ represent scalars, vectors or tensors perturbations, respectively.

Similarly, for $\beta_{ij}$, we can define the equivalent SVT modes, henceforth $svt$ modes, using rotations around any chosen vector of the invariant-basis defining a particular Bianchi model. Quite generally, it is convenient to consider $\boldsymbol{e}_3$ as a reference direction, and work with the so-called polarization basis $(\boldsymbol{e}^{(+)},\boldsymbol{e}^{(-)},\boldsymbol{e}_3)$, with $\boldsymbol{e}^{(\pm)} = (\boldsymbol{e}_1 \mp i \boldsymbol{e}_2)/\sqrt{2}$. In this basis, $\beta_{ij}$ is decomposed as
\begin{equation}\label{eq:svt}
 \beta_{ij} = \sum_{m=-2}^2 \beta_{(m)}q^{(m)}_{ij}(\boldsymbol{e}_3)\,,
\end{equation}
where the symmetric and traceless tensors $q^{(m)}_{ij}$ are the homogeneous equivalent of the tensors $Q^{(m)}_{ij}$, and given by
\begin{align}
 q^{(0)}_{ij}(\boldsymbol{e}_3) & = (-e^3_i e^3_j + \delta_{ij}/3)\,, \\
 q^{(\pm1)}_{ij}(\boldsymbol{e}_3) & = \pm e^3_{(i} e^{(\mp)}_{j)}\,, \\
 q^{(\pm2)}_{ij}(\boldsymbol{e}_3) & = -\sqrt{3/2}\,e^{(\mp)}_i e^{(\mp)}_j\,.
\end{align}

Finally, the proof of the equivalence \eqref{eq:duality} is done in two steps. First, from the linearized Einstein equations for the metrics \eqref{eq:metric-flrw} and \eqref{eq:metric-small-beta}, one looks for the wavenumber $\nu_m$ such that the SVT and $svt$ modes are dynamically equivalent. Second, for each such $\nu_m$, one verifies the equivalence of Eqs. \eqref{eq:SVT} and \eqref{eq:svt}, along with all their derivatives, at the origin of the coordinates. This leads to~\cite{Pereira:2019mpp}
\begin{equation}\label{eq:duality2}
h_{(m)}\;\,\longrightarrow\;\,\beta_{(m)}\frac{\xi_2}{\xi_m}\,
\end{equation}
in the homogeneous limit. Here, $\xi_m$ is a constant depending on the curvature radius $\ellc$ of the spatial sections (equal to $\infty$ in $\mathbb{E}^3$) and given by
\begin{equation}
\xi_m \equiv \prod_{i=1}^{|m|}\frac{k_m\ellc}{\sqrt{(\nu_m\ellc)^2- {\cal K}i^2}} \,,
\end{equation}
where
\begin{equation}
K = \pm1/\ellc^2\,,\qquad {\cal K} = K/|K|\,,
\end{equation}
and $\xi_0\equiv1$. Note also the appearance of the flat-space wavenumber $k_m$, which is identical to $\nu_m$ when ${K}=0$, and otherwise given by~\cite{Pitrou:2019ifq}
\begin{equation}\label{eq:k2_to_nu2}
\nu^2_m = k^2_m + (1+|m|)K\,.
\end{equation}
%%%%%%%%%%%%%%%%%%%%%%%%%%%%%%%%%%%%%%%%%%%%%%%
%%%%%%%%%%%%%%%%%%% Table 1 %%%%%%%%%%%%%%%%%%%
%\begin{widetext}
\begin{center}
\begin{table}
%\begin{centering}
\begin{tabular}{c c c c}
\toprule
\textbf{Type} & \textbf{$K$} & $\boldsymbol{\nu_{m}}$ & $\boldsymbol{\zeta_{\ell}^m}$ \\
\midrule
I & $=0$ & $0$  & $\delta_{\ell}^{2}$ \\
VII$_{0}$ & $=0$ & $m/\ell_{s}$ & $\left(\pm1\right)^{\ell}$ \\
IX & $>0$ & $\pm3/\ell_{c}$ & $\delta_{\ell}^{2},\; |m|=2$ only \\
V & $<0$ & $\ii/\ell_{c}$  & $(\pm1)^{m}\left(-\ii\right)^{\ell-m}\sqrt{\frac{m(m+1)}{\ell(\ell+1)}}$ \\
VII$_{h}$ & $<0$ & $\frac{m}{\ell_{s}}+\frac{\ii}{\ell_{c}}$ & $(\pm1)^{m}\!\prod_{p=m+1}^{\ell}(-\ii)\sqrt{\!\frac{(p-1)\sqrt{h}\pm \ii m}{(p+1)\sqrt{h}\mp \ii m}}$ \\
\bottomrule
\end{tabular}
%\par\end{centering}
\caption{Wavenumbers $\nu_m$ defining the limits where FLRW metric perturbations become homogeneous. Note that the limits are different for different perturbative modes $m$, and it is only trivial ($\nu_m=0$) for model I. The $\zeta^m_\ell$ coefficients result from the generalization of the plane-wave expansion to curved spaces (see \cite{Pitrou:2019ifq}), and appear explicitly in the hierarchy.}\label{tab:eigenmodes}
\end{table}
\end{center}
%\end{widetext}
%%%%%%%%%%%%%%%%%%%%%%%%%%%%%%%%%%%%%%%%%%%%%%%
%%%%%%%%%%%%%%%%%%%%%%%%%%%%%%%%%%%%%%%%%%%%%%%

The expressions for $\nu_m$ in each nearly-isotropic model is given in Table~\ref{tab:eigenmodes}. Note that it is only in the model I that the homogeneous limit corresponds to perturbations with infinite wavelength $(\nu_m=0)$ for all modes. Even for model VII$_0$, which also has flat spatial sections, the homogeneous limit will in general correspond to perturbations of finite wavelength, with the exact length depending on whether the perturbation is a scalar, vector or tensor. Notice also the appearance of a new dimensional scale ($\ells$), known as spiralling scale, which corresponds to the helicoidal isometries of some models, and which is responsible to a nontrivial multipolar structure in CMB anisotropies, as we will see. Remarkably, the homogeneous limit of hyperbolic models (V and VII$_h$) corresponds to complex wavenumbers. This means that such perturbations will have wavelengths larger than the curvature radius of the universe, and are thus supercuvature~\cite{Lyth:1995cw}.

%================================================
\section{CMB in the Line-of-sight approach}\label{sec:CMB_and_los}
%================================================

One of the main advantages of the identification between FLRW perturbations and nearly-isotropic models is the fact that observables of the latter can be computed from known expressions of the former, allowing for a unified description. In particular, the imprints of nearly-isotropic models on CMB anisotropies, which so far have been obtained using two separate codes, one for deterministic Bianchi anisotropies and another for stochastic FLRW perturbations \cite{Saadeh:2016bmp,Saadeh:2016sak}, can be unified into a single line-of-sight integrator.\footnote{See also  \cite{McEwen:2005bm} for a numerical implementation of temperature anisotropies in model VII$_h$ considering only the Sachs-Wolfe effect.} This requires two main modifications in the structure of standard codes, one related to the Boltzmann hierarchy, and another to the domain of the radial Bessel functions entering the line-of-sight integrals.

\subsection{Boltzmann-Bianchi hierarchy}\label{subsec:hierarchy}
To derive the hierarchy for Bianchi anisotropies from the standard FLRW case (see, e.g., Eqs. (33) and (34) in Ref.~\cite{Hu:1997mn}), one simply replaces the spin-multipole coupling constants $\mykp{s}{m}{\ell}$ there with a new set of constants, according to the following rule~\cite{Pereira:2019mpp}:
\begin{align}
\begin{split}
\mykp{s}{m}{\ell} & \rightarrow \mykp{s}{m}{\ell}\frac{\zeta^m_\ell}{\zeta^m_{\ell-1}}\,,\\
\mykp{s}{m}{\ell+1} & \rightarrow \mykp{s}{m}{\ell+1}\frac{\zeta^m_\ell}{\zeta^m_{\ell+1}}\,,
\end{split}
\end{align}
where
\begin{equation}
\mykp{s}{m}{\ell} = \sqrt{\frac{(\ell^2-m^2)(\ell^2-s^2)}{\ell^2}}\sqrt{\nu^2_m - {\cal K}\ell^2}\,,
\end{equation}
and where the constants $\zeta^m_\ell$, which are given in Table~\ref{tab:eigenmodes}, account for the introduction of ``pseudo plane-waves'' in curved spaces---see \cite{Pitrou:2019ifq} for details. The resulting Boltzmann-Bianchi hierarchy for the neutrinos ($\cN^{(m)}_\ell$), temperature ($\Theta^{(m)}_\ell$), and polarization $E$ and $B$ multipoles ($E^{(m)}_\ell$ and $B^{(m)}_\ell$), is given by:
\begin{widetext}
\begin{align}
\partial_\eta \cN^{(m)}_\ell & = \left[\frac{\mykp{0}{m}{\ell}}{2\ell-1}\frac{\zeta^m_\ell}{\zeta^{m}_{\ell-1}}\cN^{(m)}_{\ell-1} - \frac{\mykp{0}{m}{\ell+1}}{2\ell+3}\frac{\zeta^m_\ell}{\zeta^{m}_{\ell+1}}\cN^{(m)}_{\ell+1}\right] + \cG^{(m)}_\ell\,,\label{eq:boltzN} \\
\partial_\eta\Theta^{(m)}_\ell & = \left[\frac{\mykp{0}{m}{\ell}}{2\ell-1}\frac{\zeta^m_\ell}{\zeta^{m}_{\ell-1}}\Theta^{(m)}_{\ell - 1} - \frac{\mykp{0}{m}{\ell+1}}{2\ell+3}\frac{\zeta^m_\ell}{\zeta^m_{\ell+1}}\Theta^{(m)}_{\ell+1}\right] + \cG^{(m)}_\ell + \myC{\Theta}{m}{\ell} - \tau'\Theta^{(m)}_\ell\,,\label{eq:boltzT}  \\
\partial_\eta E^{(m)}_\ell & = \left[\frac{\mykp{2}{m}{\ell}}{2\ell-1}\frac{\zeta^m_\ell}{\zeta^{m}_{\ell-1}}E^{(m)}_{\ell - 1} - \frac{\mykp{2}{m}{\ell+1}}{2\ell+3}\frac{\zeta^m_\ell}{\zeta^m_{\ell+1}}E^{(m)}_{\ell+1}-\frac{2m\nu_m}{\ell(\ell+1)}B^{(m)}_\ell\right] + \myC{E}{(m)}{\ell} - \tau'E^{(m)}_\ell\,,\label{eq:boltzE} \\
\partial_\eta B^{(m)}_\ell & = \left[\frac{\mykp{2}{m}{\ell}}{2\ell-1}\frac{\zeta^m_\ell}{\zeta^{m}_{\ell-1}}B^{(m)}_{\ell - 1} - \frac{\mykp{2}{m}{\ell+1}}{2\ell+3}\frac{\zeta^m_\ell}{\zeta^m_{\ell+1}}B^{(m)}_{\ell+1}+\frac{2m\nu_m}{\ell(\ell+1)}E^{(m)}_\ell\right] + \myC{B}{(m)}{\ell} - \tau'B^{(m)}_\ell\,,\label{eq:boltzB}
\end{align}
\end{widetext}
where $\tau$ is the optical depth, and the coefficients $\cG^{(m)}_\ell$ and $\myC{X}{(m)}{\ell}$ ($X=\tspcolor{\Theta},E,B$) the gravitational and collisional terms, respectively. As we can see, the anisotropic dynamics have a direct impact on the free-streaming terms (inside square brackets) through the constants $\mykp{s}{m}{\ell}$ and $\zeta^m_\ell$. %The collisional and gravitational terms are formally the same as in standard FLRW perturbation theory, except that the appropriate metric perturbations $h_{(m)}$ are now replaced by $\beta_{(m)}$ through \eqref{eq:duality2}.

The system of equations \eqref{eq:boltzN}-\eqref{eq:boltzB} is not closed, and for that we also need a dynamical equation for the gravitational term $\cG^{(m)}_\ell$, as well as the expressions for the collisional terms $\myC{X}{(m)}{\ell}$. For tensor modes ($|m|=2$), the non-vanishing contributions are
\begin{align}
\cG^{(\pm 2)}_2 & = \beta'_{(\pm 2)}\,,\\
\myC{\Theta}{(\pm2)}{2} & = \tau'P^{(\pm2)}\,,\\
\myC{E}{(\pm2)}{2} & = -\tau'\sqrt{6}P^{(\pm2)}\,,
\end{align}
where
\begin{equation}
 P^{(m)} = \frac{1}{10}\left(\Theta^{(m)}_2 - \sqrt{6}E^{(m)}_2\right)
\end{equation}
accounts for the linear polarization and, in $\cG^{(\pm2)}_2$, we have used Eq.~\eqref{eq:duality2} to identify the shear with the gravitational potential. The tensor modes of the shear, in turn, satisfies
\begin{align}
\begin{split}\label{eq:beta_m2}
\beta''_{(\pm2)}+2\cH\beta'_{(\pm2)} & + (\nu^2_{\pm2}-K)\beta_{(\pm2)} =  \\
& \frac{8a^2}{5}\left(\rho_\gamma\Theta^{(2)}_2 + \rho_\nu\cN^{(2)}_2\right),
\end{split}
\end{align}
where $\rho_\gamma$ and $\rho_\nu$ are the photon and neutrino energy densities, respectively. Note that this equation follows formally from the FLRW equation for tensor perturbations through the identifications \eqref{eq:duality} and \eqref{eq:k2_to_nu2}.

For the vector modes ($|m|=1$), the non-zero terms are
\begin{align}
 \cG^{(\pm1)}_2 & = \frac{\mykp{0}{1}{2}}{3}\frac{\zeta^1_2}{\zeta^1_1}\beta'_{(\pm1)}\,,\\
 \myC{\Theta}{(\pm1)}{1} & = \tau' (v^{(\pm1)}_b+\beta'_{(\pm1)})\,, \\
 \myC{\Theta}{(\pm1)}{2} & = \tau'P^{(\pm1)}\,, \\
 \myC{E}{(\pm1)}{2} & = -\sqrt{6}\tau'P^{(\pm1)}\,.
\end{align}
Here, $v_b^{(\pm1)}$ is the (transverse) vector mode of the baryon velocity, whose dynamical equations is
\begin{equation}
 (v^{(\pm1)}_b)' + \cH v^{(\pm1)}_b = \frac{\tau'}{R}(\Theta^{(\pm1)}_1 - v^{(\pm1)}_b)\,,
\end{equation}
with $R=3{\rho}_b/4{\rho}_\gamma$. The shear equation for the vectors modes is
\begin{align}\label{eq:beta_m1}
 \beta''_{(\pm1)} & +2\cH\beta'_{(\pm1)} = \\
 & - \frac{8\sqrt{3}a^2}{5\sqrt{\nu^2_{\pm1}-4K}}\left(\rho_\gamma\Theta^{(\pm 1)}_2+\rho_\nu\cN^{(\pm1)}_2\right)\,.\nonumber
\end{align}
We stress that these equations are formally the same equations for metric perturbations $h_{(m)}$ written in \class, and that the connection with the shear is done through Eq. \eqref{eq:duality2}.

Before concluding this discussion, let us note that the vector equations above use the velocity and potential definitions given in Ref.~\cite{lewis2004observable}, which are not standard. In \anilos and \aniclass, we also provide the vector hierarchy using the conventions adopted in \class, which can be obtained by setting 'gauge: \texttt{newtonian}' in the input parameters file.\footnote{Note that the term 'newtonian' is rather misleading, as all our equations are actually implemented in the synchronous gauge.}

\subsection{Radial functions}
The Boltzmann-Bianchi system formed by Eqs.~\eqref{eq:boltzT}-\eqref{eq:boltzB} has a formal line-of-sight solution given by
\begin{align}
\frac{\Theta^{(m)}_\ell(\eta_0)}{2\ell+1} & = \int_0^{\eta_0}\!\dd\eta\, e^{-\tau}\sum_{\ell'}(\myC{\Theta}{(m)}{\ell'}+\cG^{(m)}_{\ell'})\nonumber\\
&\qquad\qquad\qquad\times\frac{\zeta^m_\ell}{\zeta^m_{\ell'}}{}_0\bar{\epsilon}_\ell^{(\ell'm)}(\chi),\label{eq:losT} \\
\frac{E^{(m)}_\ell(\eta_0)}{2\ell+1} & = \int_0^{\eta_0}\!\dd\eta\, e^{-\tau}\sum_{\ell'}\myC{E}{(m)}{\ell'}\frac{\zeta^m_\ell}{\zeta^m_{\ell'}}\,{}_2\bar{\epsilon}_\ell^{(\ell'm)}(\chi),\label{eq:losE} \\
\frac{B^{(m)}_\ell(\eta_0)}{2\ell+1} & = \int_0^{\eta_0}\!\dd\eta\, e^{-\tau}\sum_{\ell'}\myC{B}{(m)}{\ell'}\frac{\zeta^m_\ell}{\zeta^m_{\ell'}}\,{}_2\bar{\beta}_\ell^{(\ell'm)}(\chi)\label{eq:losB},
\end{align}
where $\chi=\eta_0-\eta$, and ${}_s\bar{\epsilon}^{(\ell'm)}_\ell$ and ${}_s\bar{\beta}^{(\ell'm)}_\ell$ are electric and magnetic radial functions, which also depend on $\nu_m$. Their definitions and overal properties are collected in~\cite{Pitrou:2019ifq}. The set of integral solutions also include a solution for the neutrinos, which is trivial, and thus not shown here.

These integral solutions are formally identical to those of FLRW perturbations, except for two important modifications: the appearance of the $\zeta^m_\ell$ terms, and the domain of the radial functions in the case of \emph{open} universes. The latter follows from the presence of supercurvatures modes (i.e., complex $\nu_m$ in Table~\ref{tab:eigenmodes}), which requires an analytical extension of these functions.

In order to implement this extension, we note that both the electric and magnetic radial functions can be written in terms of hyperspherical Bessel functions $\Phi_{\ell}^{\nu}$ and their derivatives~\cite{Pitrou:2019ifq}. For open universes, these functions can be written in terms of Legendre functions \cite{Harrison:1967zza, Abbott:1986ct}
\begin{equation}\label{eq:HBF}
	\Phi_{\ell}^{\nu}(\chi) = \sqrt{\dfrac{\pi N^\nu_{\ell}}{2\sinh{\chi}}} P^{-1/2 - \ell}_{-1/2 + i \nu} (\cosh{\chi}),
\end{equation}
where $P^{\alpha}_{\beta}(z)$ is defined for complex $\alpha$, $\beta$ and $|z|>1$~\cite{gradshteyn2014table}, and $N^\nu_{\ell}$ is a normalization factor. %This relation then allows us to find the hyperspherical Bessel functions for complex $\nu$.

For real and large $\nu$ and $\ell$, these functions can be efficiently computed using the algorithms described in \cite{Lesgourgues:2013bra, Tram:2013xpa}, and which are implemented in \texttt{CLASS}. Thanks to \eqref{eq:HBF}, these routines can be adapted to work with complex $\nu$. To this end, we choose the normalization factor $N^\nu_{\ell}$ that matches that used in \texttt{CLASS} \cite{Lesgourgues:2013bra}:
\begin{equation}
	N^\nu_{\ell} = \prod_{n = 1}^{\ell}(\nu^2 + n^2)
\end{equation}
so as to keep the same recurrence formulas given in  \cite{Tram:2013xpa}. Thus, in \aniclass, the generalization to complex numbers was carried by using \C's native complex library. A complex version of \texttt{CLASS}'s hyperspherical routines was introduced, with minor differences from the original ones to account for the complex domain. In \anilos, these algorithms were implemented using \texttt{Cython}, and are contained in the module \texttt{hybess.pyx}. We have compared our results against the arbitrary-precision implementation of $P^{\alpha}_\beta(z)$ found in the \python library \texttt{MPmath}~\cite{mpmath}, and found excellent agreement within the range of parameters we considered.

%------------------------------------------------
\subsection{Initial conditions}
%------------------------------------------------

An important difference between the usual SVT and the homogeneous $svt$ modes refers to their initial conditions. While the former are believed to result from some quantum random process during the early universe, the latter evolve from classical initial conditions. We now briefly discuss the initial conditions used to evolve $\beta_{(m)}$ and the baryon velocity $v_b$ in time in our numerical implementations.

In the absence of sources, the shear evolves as \cite{Pontzen:2009rx}
\begin{equation}\label{eq:beta_evolution}
 \beta''_{(m)} + 2\cH \beta'_{(m)} - \cS^{(m)}\beta_{(m)} = 0
\end{equation}
where $\cS^{(m)}$ is a constant accounting for the \tsp{(anisotropic)} spatial curvature---see \cite{Pontzen:2009rx,Pereira:2019mpp}. As remarked earlier, the solutions of the shear are classified as regular and irregular, according to their behavior at $\eta=0$. For $\cS^{(m)}=0$, $\beta_{(m)}$ is either a pure constant, or divergent at $\eta=0$, so that these solutions are not cosmologically interesting. Since this is the case in model I, this model is not implemented in \anilos or \aniclass. \tspcolor{The only exception is the tensor regular solution $\beta_{(\pm2)}=\text{constant}$ of model V. However, this solutions is not independent, but rather given by the $\ells\rightarrow\infty$ limit of the (regular) tensor solution of model VII$_h$.}

Regular and cosmologically interesting solutions can be found in the case of tensor perturbations ($|m|=2$), where
\begin{equation}
\cS^{(\pm2)} = -\nu^2_{\pm2} + K\,.
\end{equation}
They can be interpreted as superhorizon and non-stochastic gravitational waves which are frozen, but which become dynamical after entering the horizon. For example, in model VII$_h$ and in a matter-dominated universe, one finds $\beta_{(\pm2)}\propto j_1(\omega_{\pm}\eta)/(\omega_{\pm}\eta)$, where \tsp{$\omega_{\pm} = 2\ells^{-1}\sqrt{1\pm i\ells/\ellc}$}. This solution can display regular solutions which are either constant or oscillating, depending on the values of the constants $\ellc$ and $\ells$~\cite{Pontzen:2009rx}.

In the case of CMB, we set initial conditions deep in the radiation era, during the tight-coupling regime. Using $\cH=1/\eta$, it follows from \eqref{eq:beta_evolution} that the regular solution in this case is
\begin{equation}
\frac{\sin(\sqrt{-\cS^{(\pm2)}}\eta)}{\sqrt{-\cS^{(\pm2)}}\eta}\,.
\end{equation}
For modes which are in the superhorizon regime, initial conditions are chosen as
\begin{align}
 \beta_{(\pm2)}(\eta_\text{ini}) & = 1\,, \\
 \beta'_{(\pm2)}(\eta_\text{ini}) & = -(k^2_{\pm2} + 2K)\eta_\text{ini}/3\,,
\end{align}
where we have used \eqref{eq:k2_to_nu2}. \tspcolor{Note that the second linearly-independent solution of Eq.~\eqref{eq:beta_evolution} can be irregular. In the example above, there is an irregular solution given by $\cos(\sqrt{-\cS^{(\pm2)}}\eta)/(\sqrt{-\cS^{(\pm2)}}\eta)$. We stress that irregular solutions are not interesting, and thus not implemented in the codes.}

\tspcolor{In the absence of sources and for adiabatic initial conditions, vectors modes ($|m|=1$), just like scalar modes, will only present irregular or constant solutions. However, regular and non-trivial solutions can be observed in the presence of free-streaming neutrinos and isocurvature initial conditions.} For example, an initial condition (after neutrino decoupling) where photons and neutrinos have non-zero and nearly opposite velocities leads to a constant vector mode~\cite{lewis2004observable,Khalife:2024sqj}. In the homogeneous limit, this corresponds to a vector part of the shear that does not decay, but remains constant on superhorizon scales at early times; in the codes, this conditions is labeled \texttt{iso}. Another possibility (labeled \texttt{oct}) is if neutrinos only have an initial octopole moment \cite{rebhan1994kinetic}. Through the Boltzmann equation, this octopole moment induces an anisotropic stress at early times, which can also lead to a regular vector mode of the shear. In \anilos and \aniclass, we implement the tensor regular solution and both isocurvature quadrupole and octopole initial conditions for vector modes.

%================================================
\section{Anilos and Aniclass}\label{sec:anilos_and_aniclass}
%================================================
Having summarized the main modifications of standard Boltzmann codes needed to implement the CMB anisotropies from nearly-isotropic Bianchi models, we now proceed to describe our numerical implementations.

In a nutshell, \anilos and \aniclass are integrators that evolve Eqs.~\eqref{eq:losT}-\eqref{eq:losB} from a given set of cosmological parameters and initial conditions, and in the context of nearly-isotropic Bianchi models, thus providing the \emph{deterministic} CMB anisotropies at large scales resulting from these models. In both codes, the output is the set of CMB multipolar coefficients $\Theta^{(m)}_\ell$, $E^{(m)}_\ell$, and $B^{(m)}_\ell$, which can be readily converted into CMB maps using \texttt{healpy}. These coefficients should be seen as additional modulations to the existing primordial (and stochastic) CMB anisotropies.

\anilos is implemented in \python, which provides a high-level interface that simplifies the coding process and modifications by the user. Numerical solvers are offered by standard \texttt{NumPy} and \texttt{SciPy} libraries, which also handle array operations efficiently. It also relies on \texttt{Numba}, which speeds up functions by compiling them to optimized machine code, and \texttt{Cython}, used in the implementation of the radial functions. Complex types are also handled with ease, making the implementation of new initial conditions or exotic physics straightforward. However, this comes at the cost of \python's inherent performance limitations. Note that, despite being fully developed in \python, \anilos relies on \class's \python wrapper to provide common background and thermodynamical quantities. In particular, \anilos uses \class's conventions for fixing dimensionful quantities.

\begin{figure}
\centering
\includegraphics[width=\linewidth]{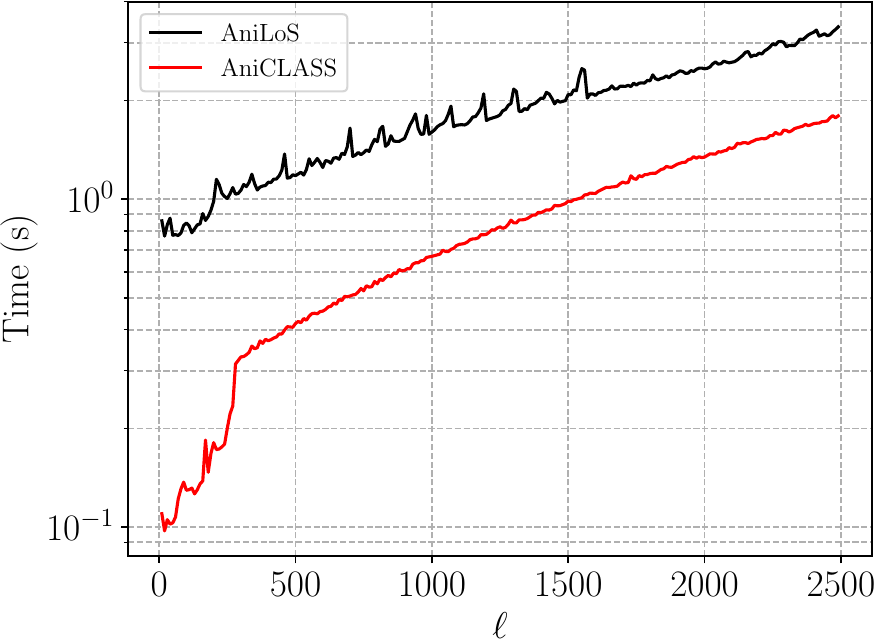}
\caption{Time performances in \aniclass and \anilos as a function of increasing multipole $\ell$, running on a Intel Core I5-8250U laptop. Overall, \aniclass is about 10 times faster than \anilos.}
\label{fig:time-comparison}
\end{figure}

In contrast, \aniclass is a low-level \C code. It is a minimal modification of \class, which already includes native and well-tested implementations of the Boltzmann hierarchy and the line-of-sight integral. However, \class is highly interconnected, meaning that localized changes often impact the entire codebase. The main departure from the original \class code is the introduction of complex numbers required to handle supercurvature modes. This is facilitated by \C's native support for complex types. While \aniclass is more difficult for end users to modify, its computational performance is unmatched, making it a powerful tool for intensive Monte Carlo simulations. Additionally, \aniclass presents a modified version of the \texttt{Python} wrapper \texttt{classy}, allowing users unfamiliar with the \texttt{C} language to compute and call the CMB multipoles directly from a \texttt{Python} interface. The time performances of both codes as a function of the output multipole $\ell$ is shown in Figure~\ref{fig:time-comparison}.

\begin{figure*}
\begin{tikzpicture}[node distance=1.0cm and -2.0cm] % reduced vertical and horizontal spacing

% Top node
\node[block, text width=7.0cm] (input) {\textbf{Input} \begin{itemize}
                                            \item Bianchi parameters ($\sqrt{h}, \Omega^0_K$)
                                            \item FLRW parameters ($H_0$, $\Omega_m$, etc.)
                                            \item Precision parameters ($\ell_\text{max}, \ell_\text{cutoff}$, $\tau_\text{reio}$)
                                            \item Pert. type (i.e., $|m|$) and vector initial conditions (isocurvature or octopole)
                                            %\item \st{Gauge (synchronous or newtonian)}
                                            \end{itemize}
};

% First row (same horizontal spacing)
\node[block, below left=of input, text width=5.2cm] (wavenumber) {\textbf{Bianchi model} $\boldsymbol{\leftrightarrow}$ \textbf{Homog. limit} \\ Computes the wavenumber $\nu_m$ from Eq.~\eqref{eq:nu_m}};
\node[block, below right=of input, text width=5.2cm] (background) {\textbf{Background} \\ Calls \class to obtain background parameters as a function of time  ($\cH$, $a$, $\chi$, $\tau$, etc.) };

% Second row (same horizontal spacing)
\node[redblock, below=of wavenumber, text width=5.2cm] (hierarchy) {\textbf{Boltzmann-Bianchi Hierarchy} \begin{itemize}
                                                                   \item Computes $\mykp{s}{m}{\ell}$ and $\zeta^{m}_\ell$
                                                                   \item Solves \eqref{eq:boltzN}-\eqref{eq:boltzB} up to $\ell_\text{max}$
                                                                  \end{itemize}
};
\node[yellowblock, below=of background, text width=5.2cm] (radialfuncs) {\textbf{Radial functions} \\ Computes electric and magnetic radial functions};

% Third row (same horizontal spacing)
\node[block, below=of hierarchy, text width=5.2cm] (sources) {\textbf{Sources} \\ Computes the gravitational and collisional sources};
\node[block, below=of radialfuncs, text width=5.2cm] (losint) {\textbf{Line-of-sight integrals} \\ Integrates Eqs.~\eqref{eq:losT}-\eqref{eq:losB}.};

% Final node
%\node[blueblock, below=1.5cm of $(sources)!0.5!(losint)$] (alms) {\textbf{CMB maps}};
\node[blueblock, below left=of losint, xshift = -0.8cm, text width=5.2cm] (alms) {\textbf{CMB multipolar coefficients} \\ $a_{\ell m}^T$, $a_{\ell m}^E$ and $a_{\ell m}^B$};

% Arrows
\draw[arrow] (input) -- (wavenumber);
\draw[arrow] (input) -- (background);

\draw[arrow] (wavenumber) -- (hierarchy);
\draw[arrow] (background) -- (radialfuncs);

\draw[arrow] (hierarchy) -- (sources);
\draw[arrow] (radialfuncs) -- (losint);

\draw[arrow] (sources) -- (losint); % new arrow
\draw[arrow] (wavenumber) -- (radialfuncs); % new arrow
\draw[arrow] (background) -- (hierarchy); % new arrow

\draw[curvedarrow] (input) to (hierarchy); % curved arrow from input to hierarchy

\draw[arrow] (losint) -- (alms);

\end{tikzpicture}
\caption{Flowchart of the \python package \anilos. The colors refer to the module where each step is found: green for \texttt{anilos.py}, red for \texttt{hierarchy.py} and yellow for \texttt{hybess.pyx}. The final output is an array containing the CMB multipolar coefficients $a_{\ell m}^T$, $a_{\ell m}^E$ and $a_{\ell m}^B$ (blue box).}\label{fig:flowchart}
\end{figure*}
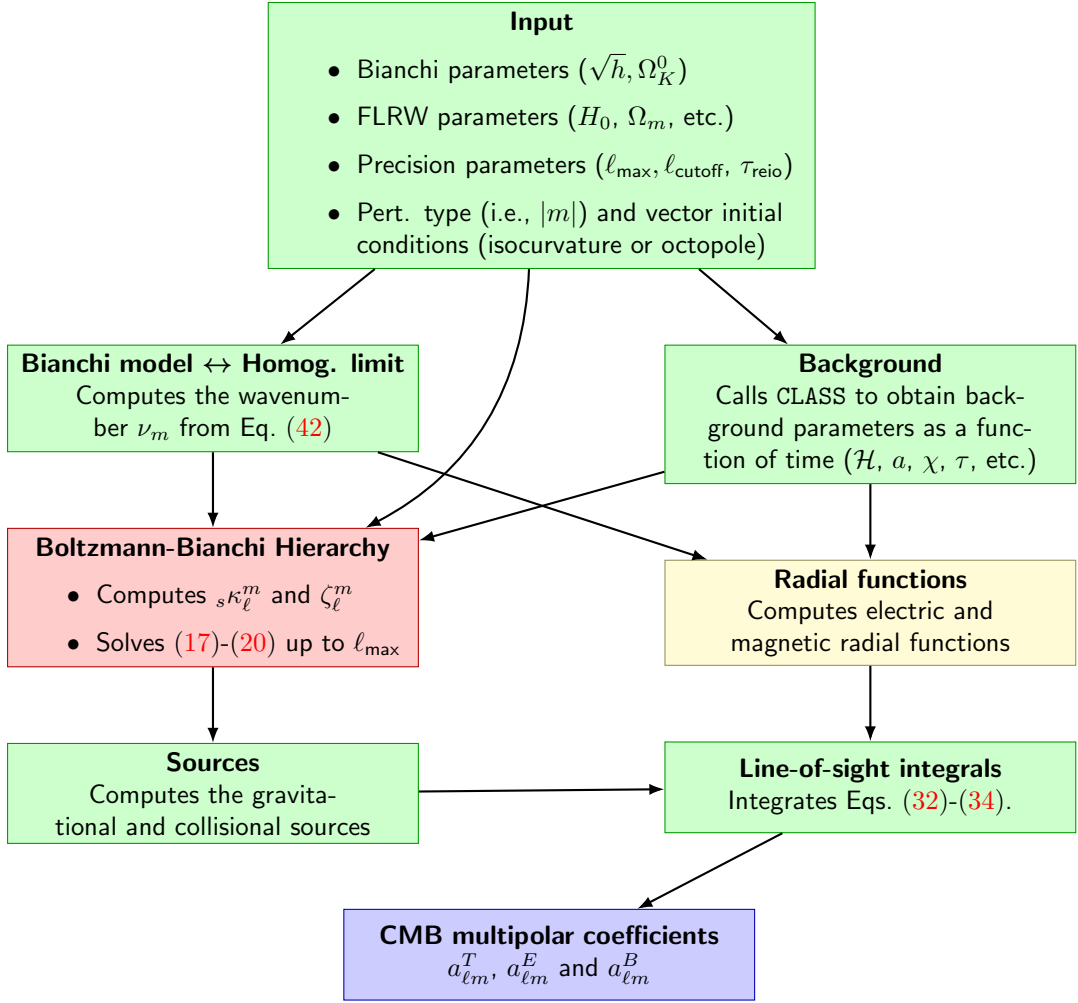

Both codes give the resulting anisotropies for vectors and/or tensor modes having the initial conditions described above. \tspcolor{Since scalar modes do not possess \emph{nontrivial} regular solutions}, they are not implemented. With the exception of model I (which does not possess regular nontrivial solutions), the codes give the resulting anisotropies from all models in Table~\ref{tab:eigenmodes}.

In practice, however, we only need to implement models VII$_h$ and IX, since models VII$_0$ and V result from the former in the limits $\ellc\rightarrow\infty$ and $\ells\rightarrow\infty$, respectively. Model IX is implemented separately, since in this case the coefficient $\zeta^{m}_\ell$ achieves a very simple form, implying that only modes with $\ell=|m|=2$ are excited, thus greatly simplifying the hierarchy and radial functions (see Eqs. (6.9) and (6.10) in~\cite{Pereira:2019mpp}). Thus, the central core of \anilos and \aniclass is devoted to solving model VII$_h$. In what follows we will detail the structure of \anilos, while stressing some important differences with \aniclass.

\anilos is organized into three main modules: \texttt{anilos.py}, \texttt{hierarchy.py}, and \texttt{hybess.pyx}. The central module, \texttt{anilos.py}, serves as the entry point to the code and is initialized with a \python dictionary that provides the key variables needed for execution. This includes:
\begin{itemize}
 \item \textbf{Bianchi parameters}: only two parameters are needed to determine a model from Table~\ref{tab:eigenmodes}. In \anilos, these are $\sqrt{h}=\ells/\ellc$ and today's curvature density, $\Omega^{0}_K$. The latter fixes $\ellc$ and, when combined with $\sqrt{h}$, also fixes $\ells$. In \aniclass, the input parameters are $m\ells^{-1}$, which is the matched Fourier mode $k$, and $\Omega^{0}_K$. Note that the Fourier mode $k$ is an input parameter in standard \texttt{CLASS}, so, to minimize the number of modifications, we opted to keep it instead of using $\sqrt{h}$. We choose $\Omega^0_K=10^{-5}$ and $\sqrt{h}=10^4$ as the default values for models VII$_0$ and V, respectively. It is possible to input values of $\Omega^0_K$ or $\sqrt{h}$ respectively smaller or higher than these values, but note that precision may be affected if the user chooses $\Omega^0_K$ too small. If $\Omega^0_K = 0 $ is chosen, then the default $\Omega^0_K=10^{-5}$ is used instead. Finally, note that model IX only requires the specification of $\Omega^{0}_K$ If the user inputs $\sqrt{h}$ or $k$, these parameters will be ignored.
 \item \textbf{FLRW parameters}: here the user can provide the usual background cosmological parameters, such as $H_0$ and $\Omega_m$. If needed, the user can also specify the conditions defining the tight-coupling regime, which is used in the computation of the Boltzmann-Bianchi hierarchy.
 \item \textbf{Precision parameters}: these refer to the maximum multipole $\ell_{\text{max}}$ used to compute the line-of-sight integral, as well as the cutoff multipole $\ell_{\text{cutoff}}$ needed to converge the Boltzmann-Bianchi hierarchy. Regarding the latter, we adopt the same truncation scheme used for stochastic linear perturbations (see Ref.~\cite{riazuelo:tel-00003366}). The user also has the ability to choose precision targets for the differential equations in \anilos. In \aniclass, the optimization options are inherited from \class.
 \item \textbf{Perturbation type and initial conditions}: the user can specify the perturbation type (i.e., the $|m|$ value) among vector, tensor, or both, as well as the initial condition for the vector modes, which can be either \texttt{iso} or \texttt{oct}. Regarding the hierarchy \eqref{eq:boltzN}-\eqref{eq:boltzT}, all multipole moments are internally chosen as zero at the initial time.
\end{itemize}

Given these input parameters, \texttt{anilos.py} determines the associated Bianchi model. More precisely, it computes the wavenumber $\nu_m$  corresponding to the homogeneous limit of FLRW perturbations. As discussed above, this falls into two cases:
\begin{equation}\label{eq:nu_m}
 \nu_m=\begin{cases}
\frac{m}{\ells} + \frac{\ii}{\ellc} & \text{if}\quad \Omega^0_K < 0\,,\\
\frac{3}{\ellc} & \text{if}\quad \Omega^0_K > 0\,.
\end{cases}
\end{equation}
This module also calls \class in order to compute the FLRW background quantities as functions of conformal time, such as ${\cal H}$, $a$, $\chi$ and $\tau$. In \anilos this is obtained by calling \class' \python wrapper \texttt{classy}, while in \aniclass these are obtained directly from the main code. The background quantities consist of arrays in conformal time, with starting and ending values determined by \class's native limits \cite{Lesgourgues:2011re}.

%starting from $\eta_\text{ini}=4\times10^{-9}\,\text{Mpc}$---which corresponds to $a_\text{ini}=2\times10^{-7}$ and $z_\text{ini}=9\times10^{13}$---up to $\eta_0=13828\,\text{Mpc}$.
%Let us mention that we use \class' native limits~\cite{Lesgourgues:2011re}.

The second module, \texttt{hierarchy.py}, contains the implementation of Boltzmann-Bianchi hierarchy for vector and tensor modes. Since the main function of this module is to provide the sources appearing in the kernels of the integrals \eqref{eq:losT}-\eqref{eq:losB}, and given that the sources are restricted to $\ell'\leq2$, it is enough to solve the hierarchy up to a small (and fixed) cutoff-multipole, and then obtain
the CMB multipoles from those integrals. Thus, when calling this module we set $\ell_\text{cutoff}=30$.

The hierarchy is implemented as ${y'=My}$, with $y$ a vector and $M$ a matrix, and integrated with \texttt{SciPy}'s \texttt{solve\_ivp} integrator using the default Runge-Kutta method. Although $M$ is sparse, implementing it as a sparse matrix proved inefficient. Instead, we found that using Numba---which compiles \python code into machine code at runtime---led to better performance. One drawback of this approach is that the first function call results in additional overhead due to compilation. However, this can be mitigated by explicitly specifying variable types, as done in \anilos.

The third module, \texttt{hybess.pyx}, is dedicated to the computation of the electric and magnetic radial functions entering the line-of-sight integrals. Since these functions will be integrated in time, they need to be sampled at a large number of time intervals, and for a many independent multipoles $(\ell,m)$. Thus, their computation is numerically intensive. As we have seen, these functions are given by combinations of hyperspherical Bessel functions and their derivatives, which obey the following recursive relations
\begin{align}
 \Phi^{\nu}_\ell & = \frac{1}{\sqrt{\nu^2 + \ell^2}}\left[(2\ell-1)\coth\chi\,\Phi^{\nu}_{\ell-1}\right. \nonumber\\
 & \qquad\qquad\quad-\left.\sqrt{\nu^2 + (\ell-1)^2}\Phi^{\nu}_{\ell-2}\right]\,, \\
 \frac{\dd\Phi^{\nu}_\ell}{\dd\chi} & = \ell\coth\chi\Phi^{\nu}_{\ell} - \sqrt{\nu^2 + (\ell+1)^2}\Phi^\nu_{\ell+1}\, ,
 \end{align}
where we have omitted the index $m$ in $\nu_m$ for simplicity. These relations provide a highly efficient computation method for real $\nu$, which is the approach adopted in \class. In our case, as we have seen, $\nu$ can take complex values. Fortunately, the recurrence relations remain valid for complex $\nu$, which means that \class's implementation can be extended to support complex $\nu$, although this is not a trivial task. In \anilos, this module is implemented using \texttt{Cython}, whereas in \aniclass these functions are obtained by a direct modification of \class' routines. In implementing these functions, both forward and backward recursion methods are employed for computing $\Phi_{\ell}^\nu$. In the forward approach, $\Phi_{\ell}^\nu$ is computed from $\Phi_{\ell - 1}^\nu$ and $\Phi_{\ell - 2}^\nu$, while in the backward approach, it is obtained from $\Phi_{\ell + 1}^\nu$ and $\Phi_{\ell + 2}^\nu$. The use of both schemes is necessary because forward recursion is numerically stable only in the region $\chi > \sinh^{-1}(\sqrt{\ell(\ell + 1) / |\nu|})$ \cite{Tram:2013xpa}. To initialize the sequence for backward recursion, we follow the procedure implemented in \class.

The integration between these three modules is illustrated in the flowchart of Figure~\ref{fig:flowchart}. Given the input parameters and the computation of auxiliary variables, \texttt{anilos.py} calls \texttt{hierarchy.py} to obtain the source terms. A call to \texttt{hybess.pyx} then provides the radial functions which, when combined with the sources, allows for a direct evaluation of the line-of-sight integrals. These integrals are schematically of the form
\begin{align}
 \frac{M^{(m)}_\ell}{2\ell+1} = \int\dd\eta \sum_{j=1}^2S^{(m)}_{j}(\eta) R^{(j m)}_\ell(\eta)
\end{align}
where $M^{(m)}_\ell$ stands for either temperature or polarization multipoles, $S^{(m)}_{j}$ represents the source terms, and $R^{(j m)}_\ell$ the radial functions. The radial functions are stored as two-dimensional arrays,
\[
 R_{\ell}^{(j m)}\rightarrow\texttt{array}\left[\begin{array}{ccc}
R_{1}^{(j m)}(\eta_{\text{ini}}) & \cdots & R_{1}^{(j m)}(\eta_{0})\\
\vdots & \ddots & \vdots\\
R_{\ell_{\text{max}}}^{(j m)}(\eta_{\text{ini}}) & \cdots & R_{\ell_{\text{max}}}^{(j m)}(\eta_{0})
\end{array}\right],
\]
and, since the source terms do not depend on $\ell$, they could be stored as one-dimensional arrays as $[S^{(m)}_j(\eta_\text{ini}),\cdots,S^{(m)}_j(\eta_0)]$. In practice, it is more efficient to use \texttt{NumPy}'s function \texttt{tile} to convert it into two-dimensional arrays,
\[
 S_{j}^{(m)}\rightarrow\texttt{array}\left[\begin{array}{ccc}
S_{j}^{(m)}(\eta_{\text{ini}}) & \cdots & S_{j}^{(m)}(\eta_{0})\\
\vdots & \ddots & \vdots\\
S_{j}^{(m)}(\eta_{\text{ini}}) & \cdots & S_{j}^{(m)}(\eta_{0})
\end{array}\right],
\]
and then vectorize their product with the radial functions. The time integrals are then implemented using \texttt{trapezoid} from \texttt{SciPy}, which leads to the multipoles $\Theta^{(m)}_\ell$, $E^{(m)}_\ell$, and $B^{(m)}_\ell$ for $\ell=0,\cdots,\ell_\text{max}$ and $|m|\leq2$. Finally, in order to connect these multipoles to the more commonly adopted multipoles accounting for observed directional anisotropies, (i.e. $a_{\ell m}^T$, $a_{\ell m}^E$, $a_{\ell m}^B$) we use \cite{Pereira:2019mpp}
%\begin{equation}
% a_{\ell m}^T = (-\ii)^{\ell}\sqrt{\frac{4\pi}{2\ell+1}}\Theta^{(m)}_\ell\,,
%\end{equation}
\begin{equation}
	a_{\ell m}^T = \ii^{\ell}\sqrt{\frac{4\pi}{2\ell+1}}\Theta^{(m)}_\ell\, ,
\end{equation}
with similar formulas for the polarization multipoles (with an additional $-1$ factor for the $B$ mode). These are the main output of our codes. The coefficients of both the total angular momentum basis and the usual spherical harmonic basis are available in \aniclass (that is, the user can choose to output $a^T_{\ell m}$ or $\Theta^{(m)}_{\ell}$, for example). In \anilos, only the former is implemented.

Note that, since nearly anisotropic models can only excite coefficients with $|m|=1,2$, many of the resulting coefficients will be zero. Thus, both in \anilos and \aniclass, the user can choose between a dense output (i.e., with only non-zero coefficients) or the full output in \texttt{healpy} format (a Jupyter notebook with examples can be found in \cite{gitjoao}). However, in \aniclass, the full output is not available in the \python wrapper. Another important aspect of both codes is that the equations implemented are specialized to the (fixed) polariation basis $(\boldsymbol{e}^{(+)},\boldsymbol{e}^{(-)},\boldsymbol{e}_3)$. When doing forecasts or real-data search for anisotropies, one should also allow for three Euler angles corresponding to the orientation of the (unknown) axes of anisotropies. In practice, this can be done by rotating the output $a_{\ell m}$s with \texttt{HEALPix}/\texttt{healpy} rotation routines.

%================================================
\section{Examples}\label{sec:examples}
%================================================

\begin{figure*}
\centering
\includegraphics[width=0.5\linewidth]{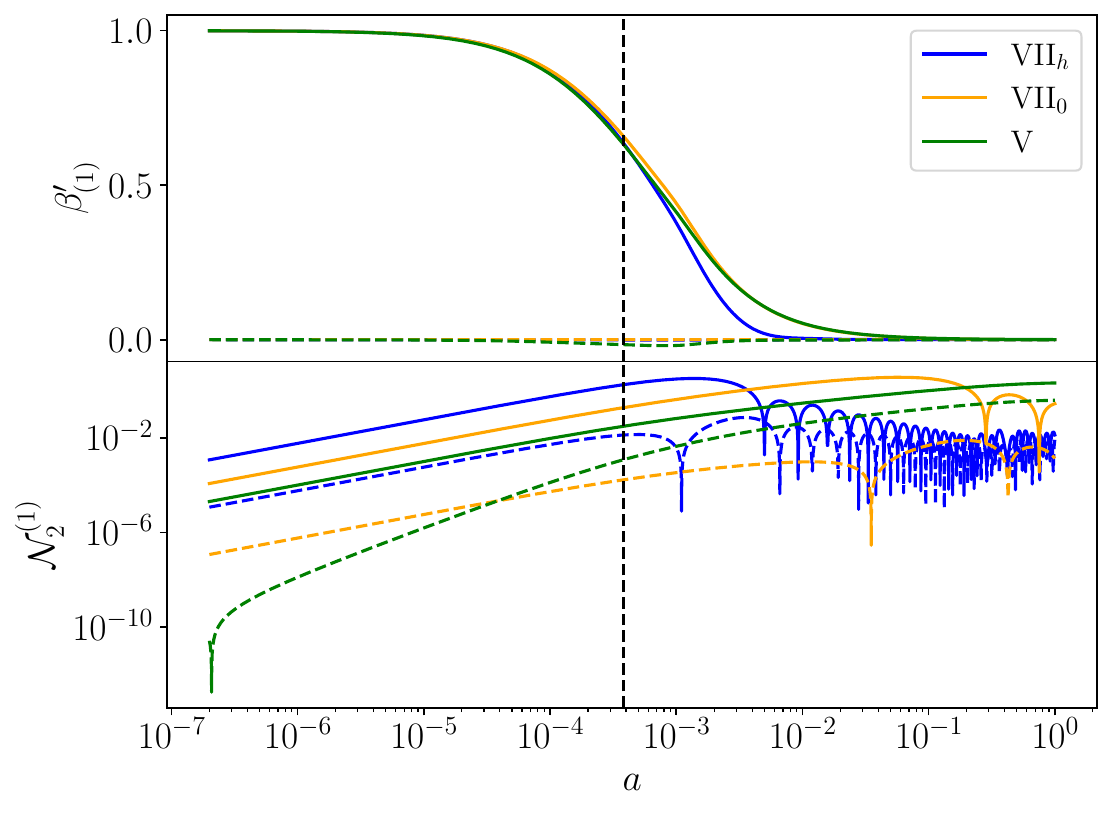}\,\includegraphics[width=0.5\linewidth]{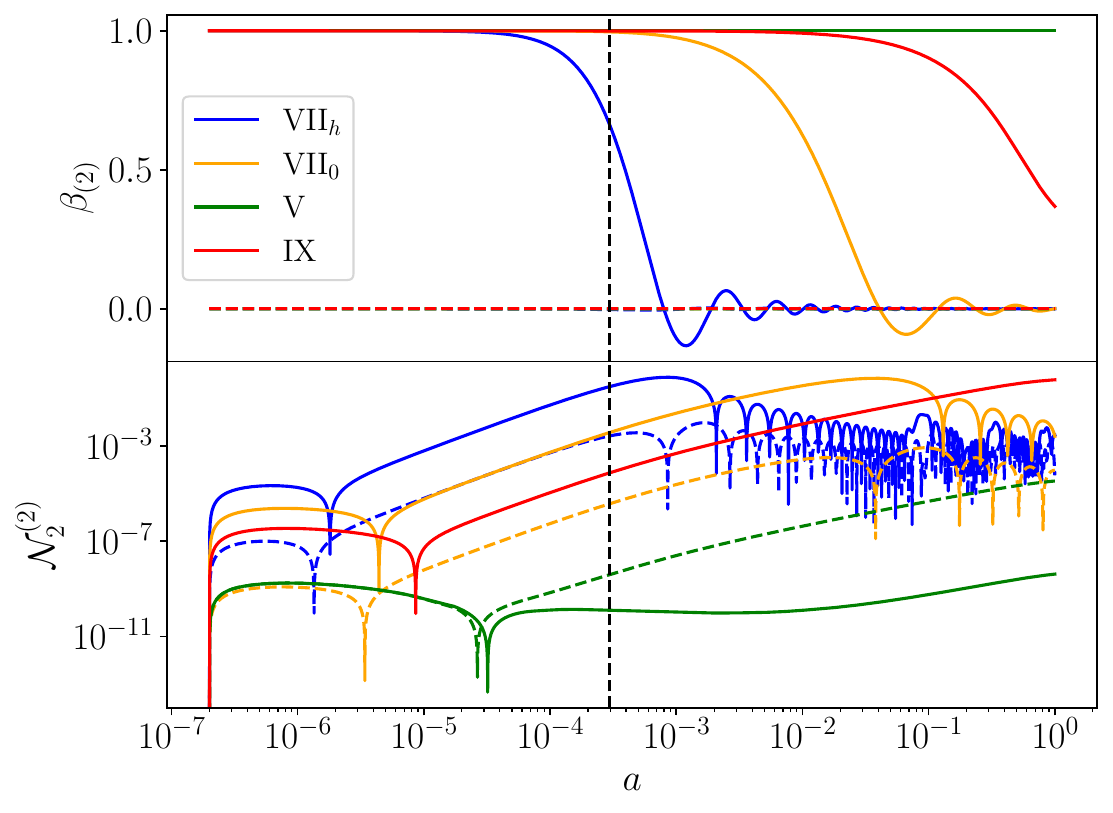}
\caption{\tsp{Left: evolution of $\beta'_{(1)}$ and the neutrino quadrupole $\cN^{(1)}_2$ for vector modes. Right: evolution of $\beta_{(2)}$ and the neutrino quadrupole $\cN^{(2)}_2$ for tensor modes. Both panels show the real (continuous) and imaginary (dashed) parts of the solution.} For this plot we have used: $\Omega_K^0 = 0.1$ (VII$_h$ and V), $10^{-5}$ (VII$_0$), and $- 0.1$ (IX); $\sqrt{h}= 0.01$ (VII$_h$), $10^4$ (V), and $10^{-3}$ (VII$_0$). For the vector modes, the neutrino isocurvature quadrupole initial condition was used. Recall that model IX has no vector modes, and is thus shown only on the right panel.}
\label{fig:beta}
\end{figure*}

\begin{figure*}
\centering
% First row
\subfloat[Regular tensor ($T$)]{%
  \includegraphics[width=0.32\linewidth]{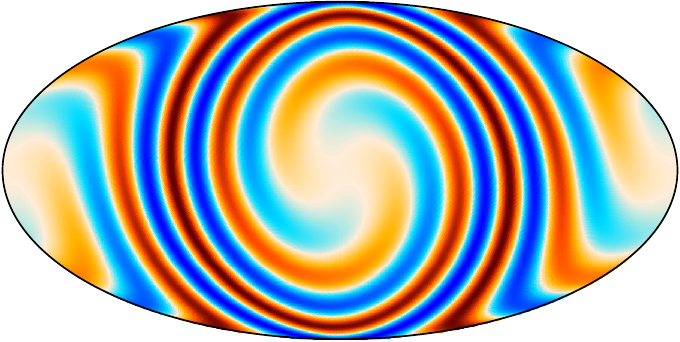}}\,
\subfloat[Regular tensor ($E$)]{%
  \includegraphics[width=0.32\linewidth]{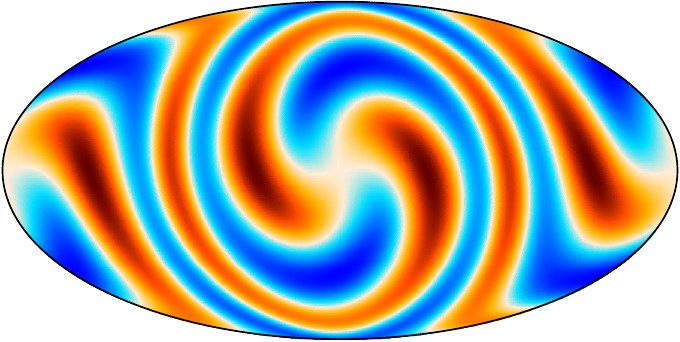}}\,
\subfloat[Regular tensor ($B$)]{%
  \includegraphics[width=0.32\linewidth]{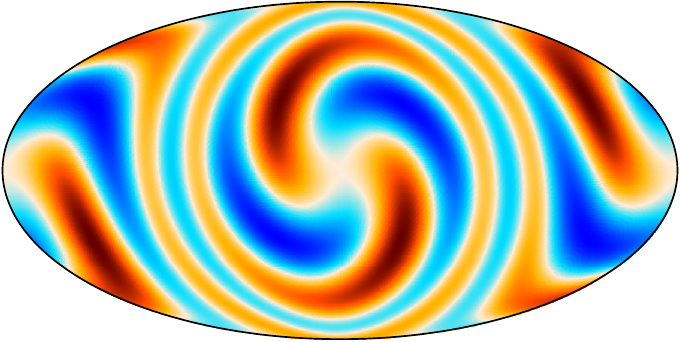}}%

% Second row
\subfloat[Vector isocurvature ($T$)]{%
  \includegraphics[width=0.32\linewidth]{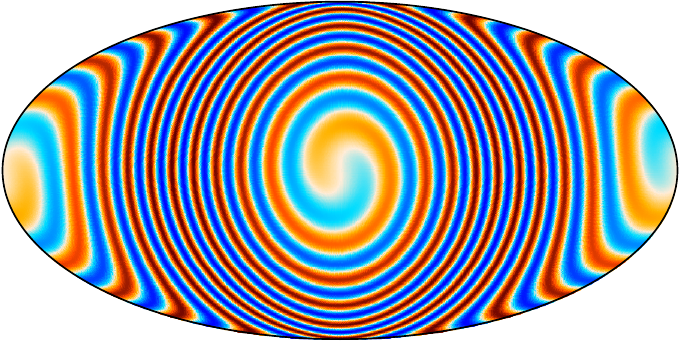}}\,
\subfloat[Vector isocurvature ($E$)]{%
  \includegraphics[width=0.32\linewidth]{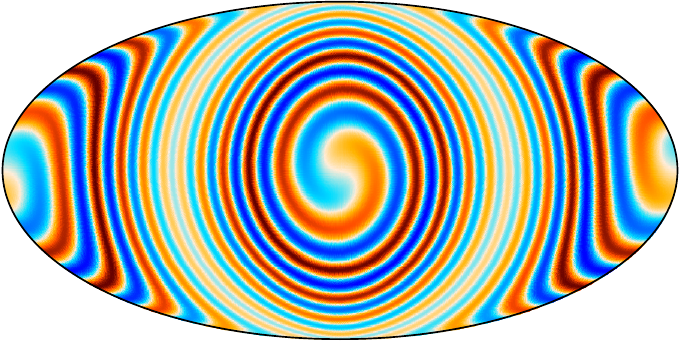}}\,
\subfloat[Vector isocurvature ($B$)]{%
  \includegraphics[width=0.32\linewidth]{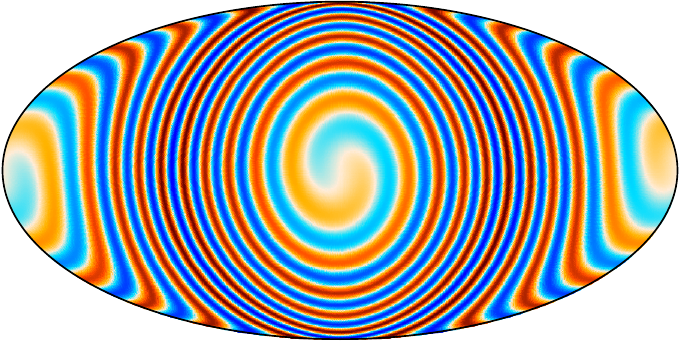}}%

% Third row
\subfloat[Vector octopole ($T$)]{%
  \includegraphics[width=0.32\linewidth]{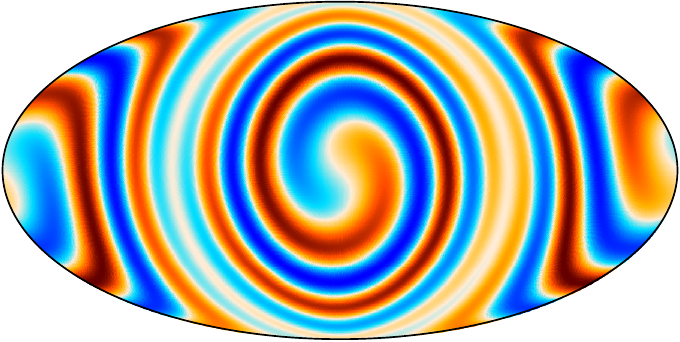}}\,
\subfloat[Vector octopole ($E$)]{%
  \includegraphics[width=0.32\linewidth]{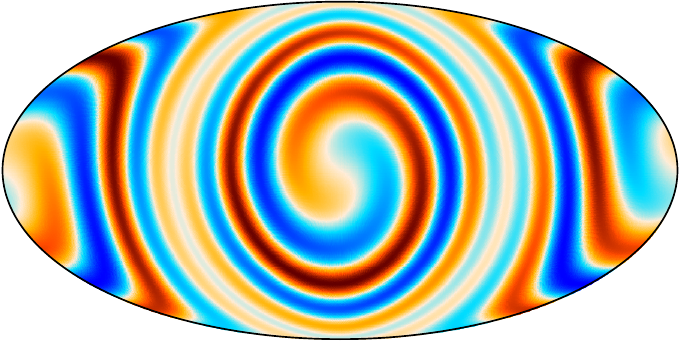}}\,
\subfloat[Vector octopole ($B$)]{%
\includegraphics[width=0.32\linewidth]{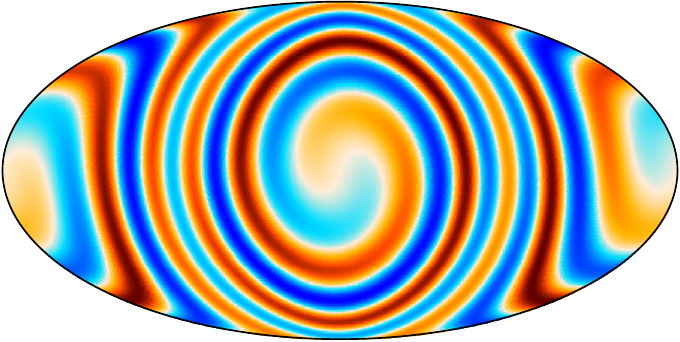}}%
\captionsetup{justification=justified,singlelinecheck=false}
\caption{CMB anisotropies in model VII$_0$, with three different initial conditions: regular tensor modes (top row), isocurvature vector modes (middle row) and octopole vector modes (botton row). The cosmological parameters used in this plot are $\Omega^0_m=0.31$, $\Omega^0_\Lambda=0.69$, $\Omega_K=10^{-5}$ and $\ells=884\,\text{Mpc}\approx20\% H^{-1}_0$.}
\label{fig:alms-VII0}
\end{figure*}

\begin{figure*}
\centering
% First row
\subfloat[Regular tensor ($T$)]{%
  \includegraphics[width=0.32\linewidth]{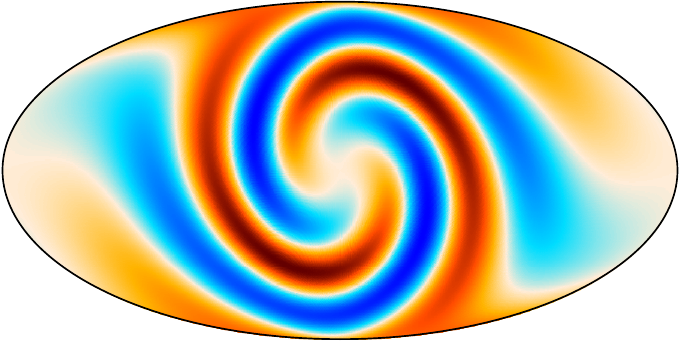}}\,
\subfloat[Regular tensor ($E$)]{%
  \includegraphics[width=0.32\linewidth]{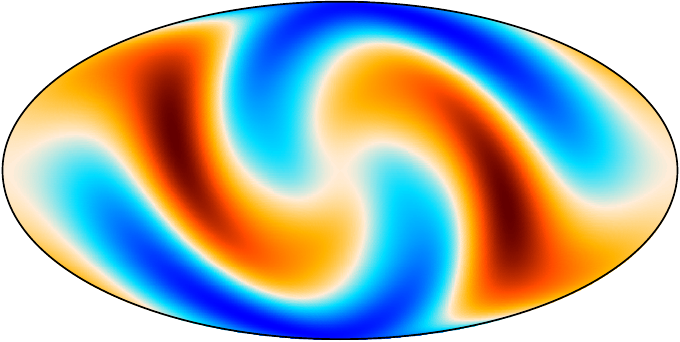}}\,
\subfloat[Regular tensor ($B$)]{%
  \includegraphics[width=0.32\linewidth]{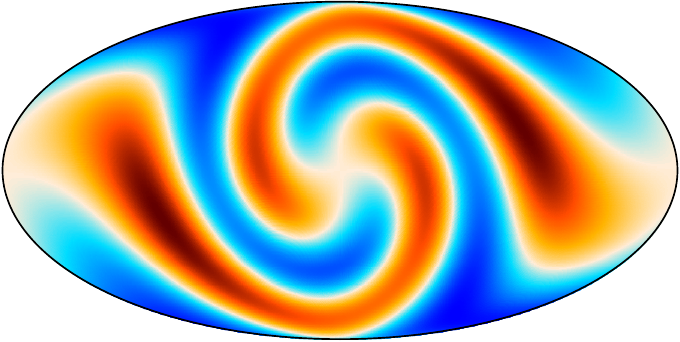}}%

% Second row
\subfloat[Vector isocurvature ($T$)]{%
  \includegraphics[width=0.32\linewidth]{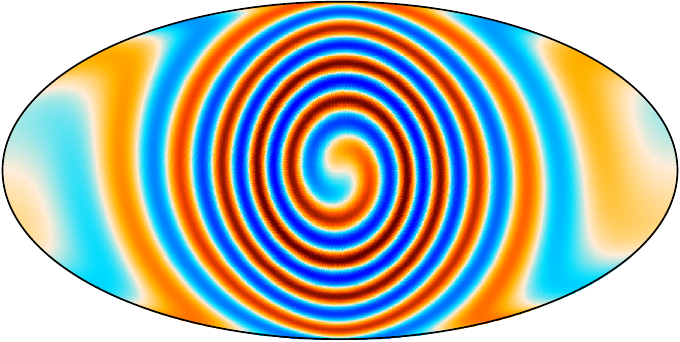}}\,
\subfloat[Vector isocurvature ($E$)]{%
  \includegraphics[width=0.32\linewidth]{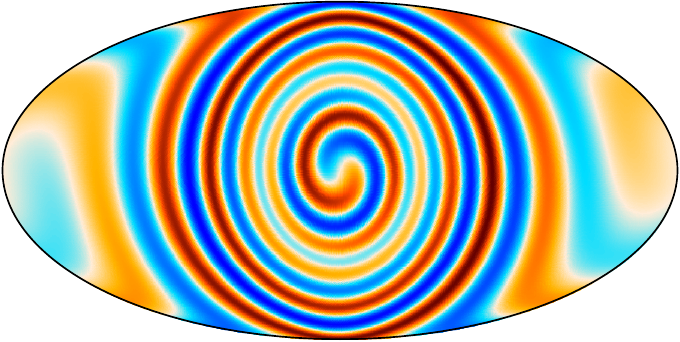}}\,
\subfloat[Vector isocurvature ($B$)]{%
  \includegraphics[width=0.32\linewidth]{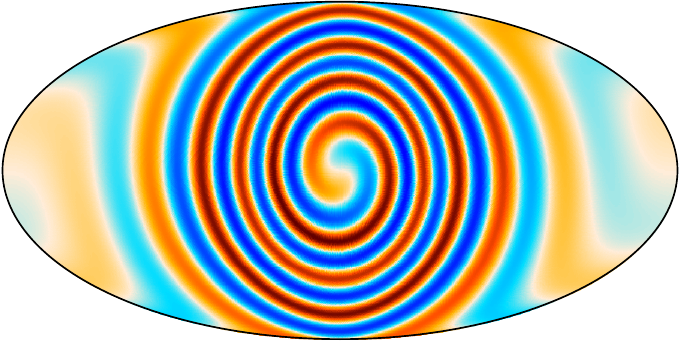}}%

% Third row
\subfloat[Vector octopole ($T$)]{%
  \includegraphics[width=0.32\linewidth]{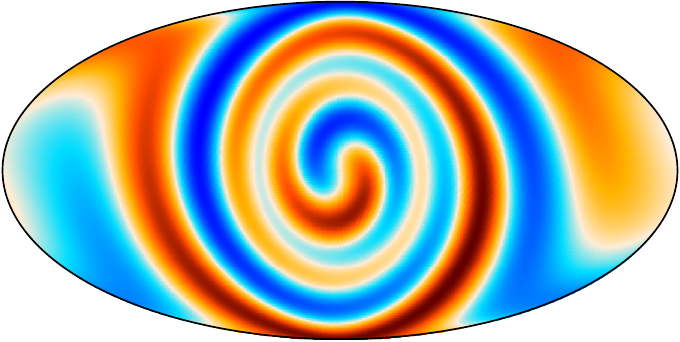}}\,
\subfloat[Vector octopole ($E$)]{%
  \includegraphics[width=0.32\linewidth]{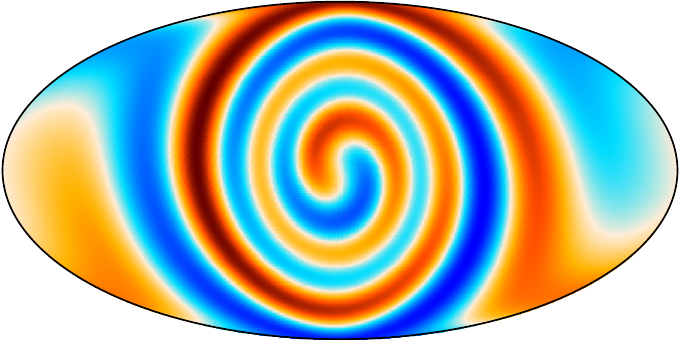}}\,
\subfloat[Vector octopole ($B$)]{%
\includegraphics[width=0.32\linewidth]{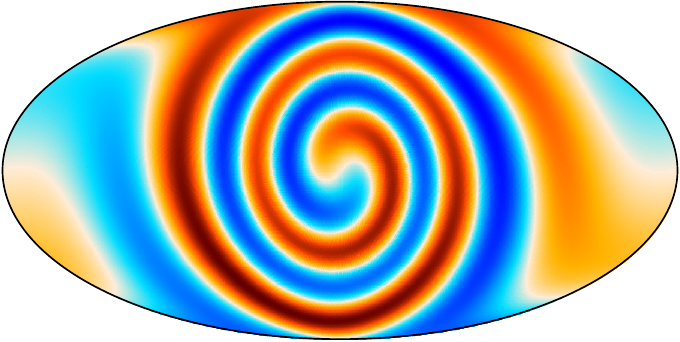}}%
\captionsetup{justification=justified,singlelinecheck=false}
\caption{Same as Figure~\ref{fig:alms-VII0} but for model VII$_h$. For this plot we have used $\Omega^0_m=0.3$, $\Omega^0_\Lambda=0.6$, $\Omega_K^0 = 0.1$, and $\ells= 800$ Mpc.}
\label{fig:alms-VIIh}
\end{figure*}

\begin{figure*}
\centering
% First row
\subfloat[Regular tensor ($T$)]{%
  \includegraphics[width=0.32\linewidth]{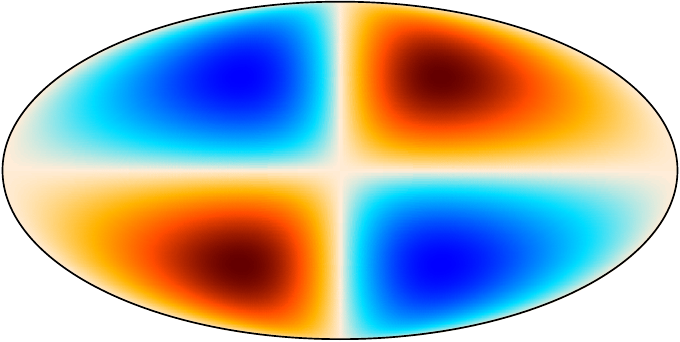}}\,
\subfloat[Regular tensor ($E$)]{%
  \includegraphics[width=0.32\linewidth]{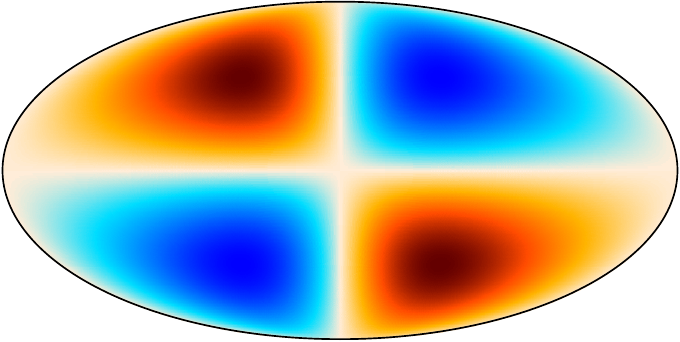}}\,
\subfloat[Regular tensor ($B$)]{%
  \includegraphics[width=0.32\linewidth]{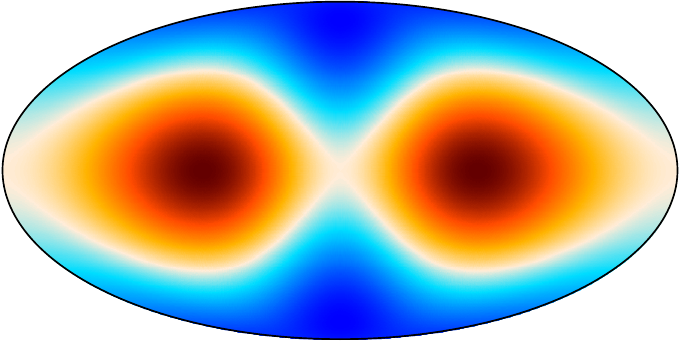}}%

% Second row
\subfloat[Isocurvature ($T$)]{%
  \includegraphics[width=0.32\linewidth]{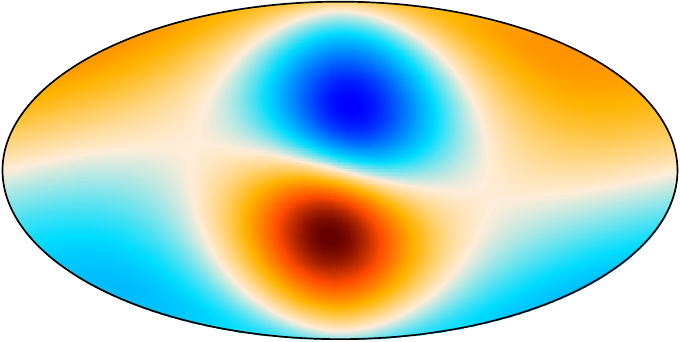}}\,
\subfloat[Isocurvature ($E$)]{%
  \includegraphics[width=0.32\linewidth]{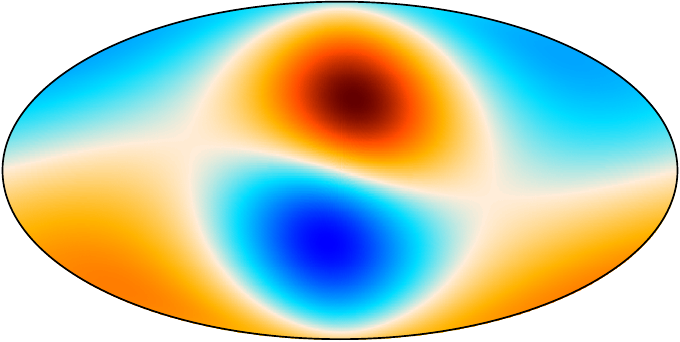}}\,
\subfloat[Isocurvature ($B$)]{%
  \includegraphics[width=0.32\linewidth]{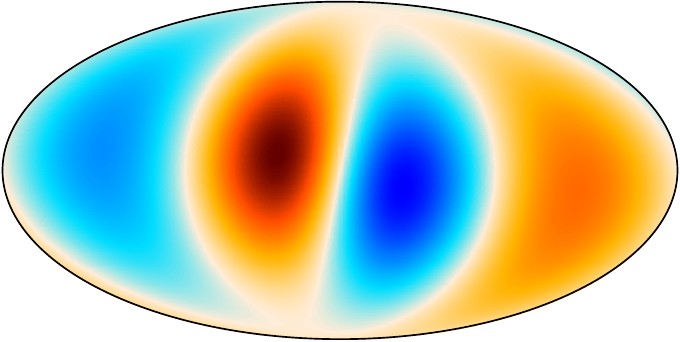}}%

% Third row
\subfloat[Octopole ($T$)]{%
  \includegraphics[width=0.32\linewidth]{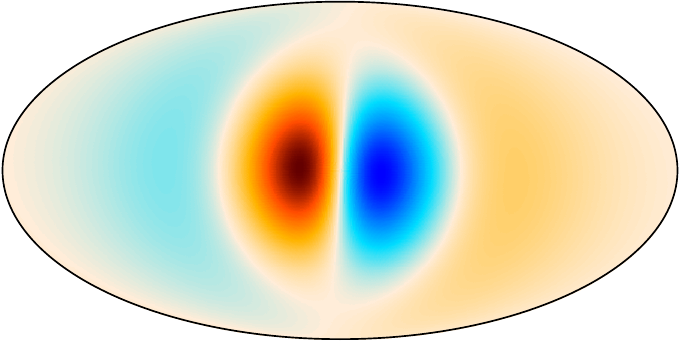}}\,
\subfloat[Octopole ($E$)]{%
  \includegraphics[width=0.32\linewidth]{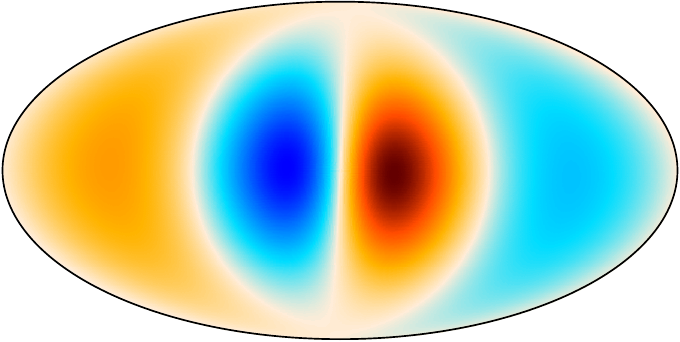}}\,
\subfloat[Octopole ($B$)]{%
\includegraphics[width=0.32\linewidth]{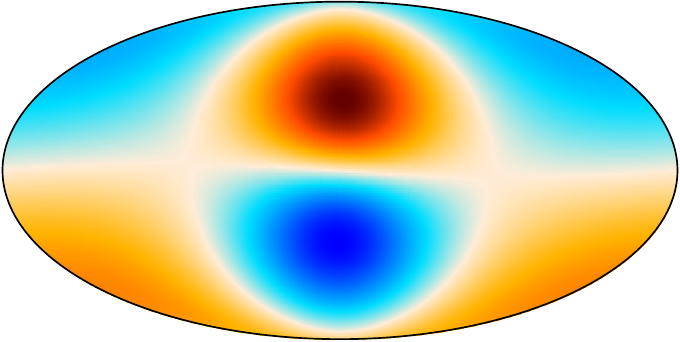}}%
\captionsetup{justification=justified,singlelinecheck=false}
\caption{Same as Figure~\ref{fig:alms-VII0} but for model V. For this plot we have used $\Omega^0_m=0.3$, $\Omega^0_\Lambda=0.6$, $\Omega_K^0 = 0.1$, and $\ells= 10^9$ Mpc, or $\sqrt{h} = 10^5$.}
\label{fig:alms-V}
\end{figure*}

\begin{figure*}
\centering
% First row
\subfloat[Regular tensor ($T$)]{%
  \includegraphics[width=0.32\linewidth]{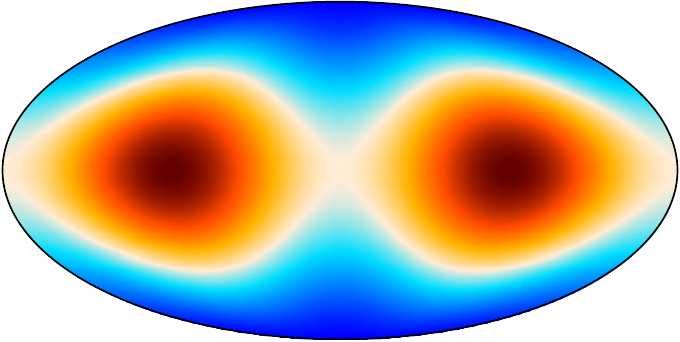}}\,
\subfloat[Regular tensor ($E$)]{%
  \includegraphics[width=0.32\linewidth]{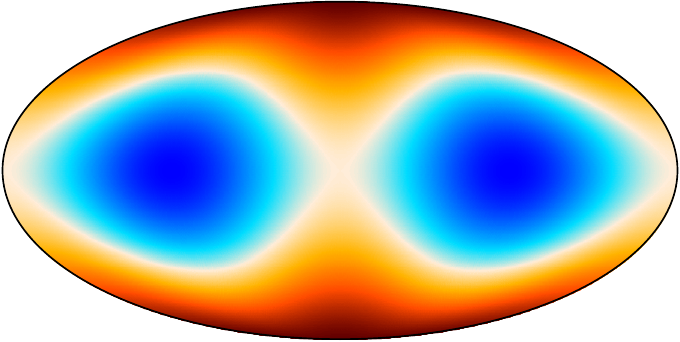}}\,
\subfloat[Regular tensor ($B$)]{%
  \includegraphics[width=0.32\linewidth]{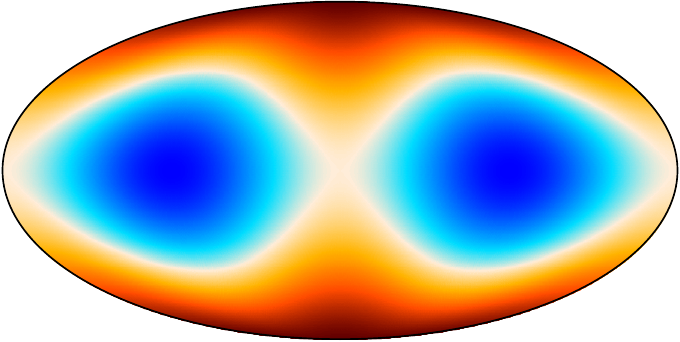}}%
\captionsetup{justification=justified,singlelinecheck=false}
\caption{CMB anisotropies for model IX. Note that in this case we only have a quadrupolar modulation. For this plot we have used $\Omega_m^0 = 0.3$, $\Omega_\Lambda^0 = 0.8$, and $\Omega^0_K=-0.01$.}
\label{fig:alms-IX}
\end{figure*}

Let us now illustrate our codes with a few examples. Figure \ref{fig:beta} illustrates the evolution of vector and tensor modes of the shear and neutrino quadrupole. Note how the tensor component of the shear is initially constant despite the small values of anisotropic stress (right panel). Conversely, the vector component of the shear is only initially constant because of the intensity of the anisotropic stress. \tspcolor{The right panel of Figure \ref{fig:beta} also displays the trivial tensor solution for model V, normalized as $\beta_{(\pm2)}\equiv1$, and which follows as a limit of the tensor solution of model VII$_h$, as explained above.}
Figures \ref{fig:alms-VII0} to \ref{fig:alms-IX} showcase the expected temperature and polarization CMB anisotropies in all nearly-isotropic models (except model I), both for regular tensor modes and the two possible regular vector modes we considered. In all these figures, the $z$-axis was chosen to point along the normal to the screen, so as to help with the visualization of the anisotropic features.

Models VII$_0$ and VII$_h$ are the only cases where spatial translations are accompanied by corkscrew rotations. Consequently, these models display a clear spiraling pattern seen in Figures~\ref{fig:alms-VII0} and \ref{fig:alms-VIIh} (see also \cite{Pontzen:2009rx}). Notice how the spiralling patterns for the vector modes are stronger than those found in the tensor modes (2nd and 3rd rows in both figures). This is because the typical spiral scale is given by $2\pi\ells/|m|$ in these models.\footnote{\tspcolor{The best available constraints on this scale come from the Wilkinson Microwave Anisotropy Probe (WMAP) data, and give $x\equiv\ells/H_0^{-1}\approx1.2$ \cite{Saadeh:2016bmp}. However, the posterior distribution of $x$ is very broad (see Figure 5 of Ref. \cite{Saadeh:2016bmp}), so that current constraints on $\ells$ are weak.}} However, from the observational point of view, the most important feature of these models lies in the multipolar structure, which goes beyond a simple  $\ell=2$. This follows from the non-trivial structure of the $\zeta^m_\ell$ coefficients entering the hierarchy (see Table~\ref{tab:eigenmodes}), which in turn reflects the non-trivial kinematics of light in these models. This is in sharp contrast to the anisotropies of models V and IX shown in Figures~\ref{fig:alms-V} and \ref{fig:alms-IX}, which can only display quadrupolar anisotropies. Models V and VII$_h$ also display a focusing of the quadrupole pattern along the $z$-axis, which is a typical signature of hyperbolic spaces~\cite{Barrow:1997vu}.

\begin{figure*}
\centering
\includegraphics[width=0.5\linewidth]{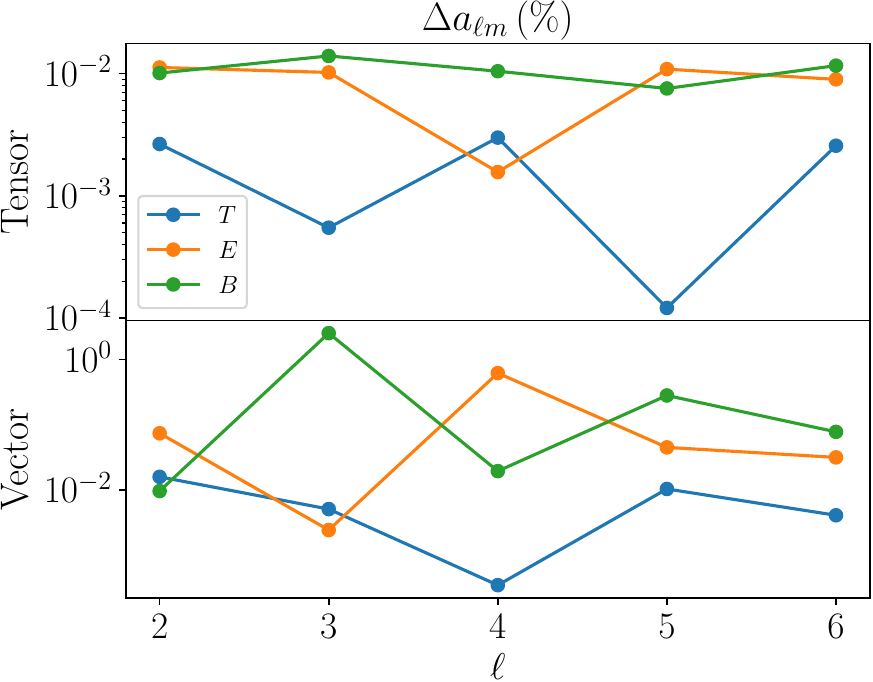}\,\includegraphics[width=0.47\linewidth]{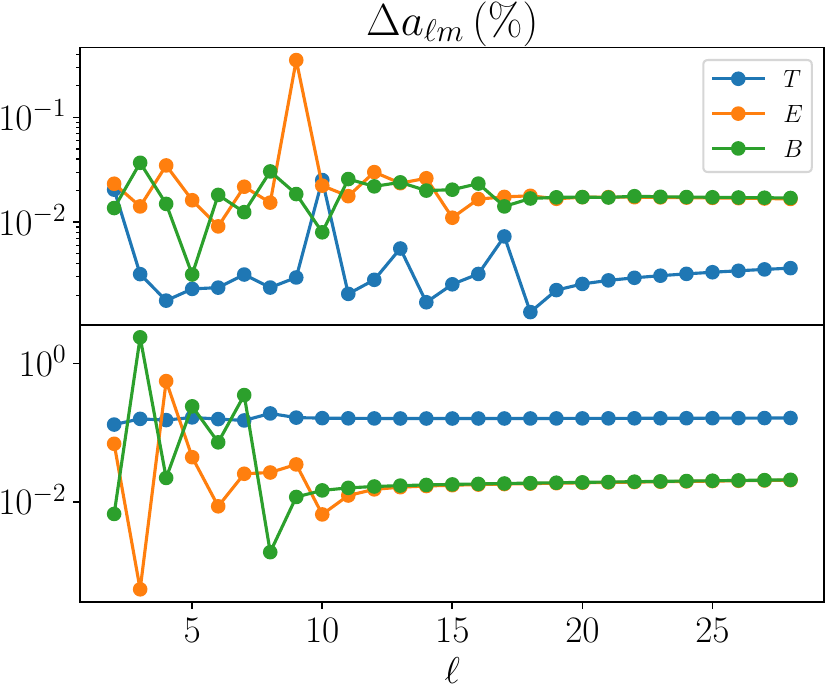}
\caption{Relative differences between the main outputs of \anilos and \aniclass, both for vector (lower panel) and tensor modes (upper panel). \textit{Left}: relative differences between the multipolar coefficients resulting from the hierarchy \eqref{eq:boltzN}-\eqref{eq:boltzB} up to $\ell=6$, which is the cutoff used for integrating the source terms. \textit{Right}: relative differences between the full harmonic coefficients for temperature and polarization. For this plot we have used $\Omega_K^0 = 10^{-5}$ and $\sqrt{h}= 8 \times 10^{-4}$, which corresponds to $\ell_s \sim 1000$ Mpc, and isocurvature initial conditions for the vector modes.}\label{fig:relative-diff}
\end{figure*}

Finally, let us comment on the numerical differences between \anilos and \aniclass. Figure~\ref{fig:relative-diff} shows a comparison of the two main outputs: i) solutions of the Boltzmann-Bianchi hierarchy for the source terms and ii) the integration of these sources along the line of sight. As we can see, for cosmological parameters not too far from observational limits, relative differences no larger than $2\%$ are observed at the lowest multipoles, but are overall smaller than $\sim0.1\%$ for $\ell>10$.
Typically, extreme values of $\Omega^0_K$, $\sqrt{h}$, and $\ells$ will lead to numerical instabilities, and thus larger relative differences. For example, in model VII$_0$, a spiralling scale of $\ells\sim 0.1\text{Gpc}$, thus much smaller than the Hubble radius, will lead to $\sim 10\%$ relative differences between the outputs of the two codes. This isn't surprising since \anilos uses \texttt{Scipy}'s default settings for accuracy, and was not particularly optimized for dealing with stiff systems. This aligns with the philosophy of \anilos, of being user-friendly and easy to modify. Overall, when targetting precision, \aniclass should be preferred over \anilos, since it relies on \class infrastructure, whose numerical precision is under control~\cite{Lesgourgues:2011re}.

%==========================
\section{Conclusions}\label{sec:conclusions}
%==========================

\tspcolor{Upcoming cosmological observations, particularly those probing near-horizon scales --- such as LiteBIRD \cite{LiteBIRD:2022cnt} --- have the potential to significantly improve our understanding of the Universe, including more robust constraints on its geometry at horizon scales.} This task will not only benefit from new tools allowing an efficient comparison between data and models, but also an integration of existing tools. In this work, we have introduced two numerical softwares, \anilos and \aniclass, allowing for a unified treatment of stochastic and deterministic CMB anisotropies. In particular, when the deterministic component arises from nearly-isotropic cosmological models, it was shown in \cite{Pereira:2019mpp} that the resulting anisotropies can be obtained from standard line-of-sight CMB codes with minimal modifications.

Although both softwares perform the same tasks, they have different use cases. \anilos is a user-friendly \python package allowing for easy inclusion of new initial conditions and exotic physics, whereas \aniclass, an extension of \class, targets at efficiency and intensive Monte Carlo simulations. Both codes are free to download and use at~\cite{gitjoao}.

\anilos and \aniclass can be extended in many interesting ways beyond the inclusion of new (regular) initial conditions. One obvious example is the inclusion of massive neutrinos, which would directly affect the tensor modes of the shear.
From a more phenomenological perspective, one can consider that dark energy develops anisotropic pressure at late times~\cite{Koivisto:2008ig,Pereira:2015jya,BeltranAlmeida:2021ywl}, which could act as a direct source to the right-hand side of Eq.~\eqref{eq:beta_evolution}. In this case, regular solutions can develop even if $\cS^{(m)}=0$, and scalar modes of the shear ($m=0$) can develop interesting signatures.

\tspcolor{The possibility of including non-trivial initial conditions may be relevant for investigating the origin of low-$\ell$ CMB anomalies. As illustrated in the figures of Section \ref{sec:examples}, a quadrupolar modulation of temperature and polarization anisotropies is an ubiquitous signature of the shear. At the same time, the low value of the CMB quadrupole \cite{Bennett:1996ce,WMAP:2003ivt,Planck:2013lks} and its alignment with the octopole \cite{deOliveira-Costa:2003utu,Schwarz:2004gk} are one of the most robust statistical anomalies, the latter persisting even after correction for the frequency-dependent kinetic Doppler quadrupole \cite{Notari:2015kla}. Although the amplitude of the shear is tightly constrained by Planck \cite{Saadeh:2016sak}, many of these anomalies could be associated with off-diagonal correlations between $a_{\ell m}$'s arising from a possible breakdown of statistical isotropy, which could in turn evade such constraints. Whether these anomalies are physical and, if so, of a deterministic or stochastic origin, is yet not known.}

More generally, this work paves the way for a unified framework to describe the imprints of nearly-isotropic cosmologies on all cosmological observables described as past light-cone integrals over source fields (e.g., weak lensing shear, galaxy number counts, redshift and position drifts, etc.). Since so far only the effect of scalar modes is implemented for these observables, one would need first to implement the effect of vector and tensor perturbations on them. As in the case of the CMB, applying this method to a given observable would then involve two key steps: properly incorporating the homogeneous limit and the pseudo plane-wave coefficients $\zeta^m_\ell$ into the corresponding equations of motion, and analytically extending the radial Bessel functions that appear in the light-cone integrals. Given the constraining power of current and upcoming large-scale structure surveys on late-time anisotropies, this direction holds significant promise and will be pursued in future work.

\newpage

\begin{acknowledgements}
We thank Thomas Tram for useful discussions on the analytical extension of hyperspherical Bessel functions, and Nils Schöneberg and Julien Lesgourgues for their support with CLASS.  J.G.V thanks Brazilian agencies CAPES and CNPq for financial support. T.S.P. is supported by FAPERJ (grant E26/204.633/2024), CNPq (grant 312869/2021-5) and Fundação Araucária (NAPI de Fenômenos Extremos do Universo, Grant No. 347/2024 PDI). This work made extensive use of the following numerical packages: \texttt{NumPy} \cite{harris2020array}, \texttt{SciPy} \cite{2020SciPy-NMeth}, \texttt{mpmath} \cite{mpmath}, \texttt{HEALPix} \cite{gorski2005healpix}, \texttt{healpy} \cite{zonca2019healpy}, \texttt{Numba} \cite{lam2015numba}, \texttt{Cython} \cite{behnel2011cython}, and \texttt{Matplotlib} \cite{Hunter:2007}.
\end{acknowledgements}

\section{References}
\bibliographystyle{h-physrev4}
\bibliography{anilos}

\end{document}